\documentclass[lettersize,journal]{IEEEtran}
\usepackage{amsmath,amsfonts}
\usepackage{algorithmic}
\usepackage{algorithm}
\usepackage{array}
\usepackage{subfigure}
\usepackage{textcomp}
\usepackage{stfloats}
\usepackage{url}
\usepackage{verbatim}
\usepackage{graphicx}
\usepackage{cite}
\usepackage[dvipsnames]{xcolor}
\usepackage{gensymb}
\usepackage{makecell}
\usepackage{enumitem}
\definecolor{bluerevision}{RGB}{0,112,192}

\makeatletter
 \let\old@ps@headings\ps@headings
 \let\old@ps@IEEEtitlepagestyle\ps@IEEEtitlepagestyle
 \def\confheader#1{%
 \def\ps@headings{%
 \old@ps@headings%
 \def\@oddhead{\strut\hfill#1\hfill\strut}%
 \def\@evenhead{\strut\hfill#1\hfill\strut}%
 }%
 \def\ps@IEEEtitlepagestyle{%
 \old@ps@IEEEtitlepagestyle%
 \def\@oddhead{\strut\hfill#1\hfill\strut}%
 \def\@evenhead{\strut\hfill#1\hfill\strut}%
 }%
 \ps@headings%
 }
 \makeatother
\confheader{%
This work has been accepted for publication in \textit{IEEE Transactions on Vehicular Technology}. DOI: 10.1109/TVT.2024.3393533
}

\begin{document}

\title{FR2 5G Networks for Industrial Scenarios:\\Experimental Characterization and Beam Management Procedures in Operational Conditions}

\author{Alejandro Ramírez-Arroyo, Melisa López, Ignacio Rodríguez, Troels B. Sørensen, \linebreak Samantha Caporal del Barrio, Pablo Padilla, Juan F. Valenzuela-Valdés, Preben Mogensen
\thanks{This work has been supported by grant TED2021-129938B-I00 funded by MCIN/AEI/10.13039/501100011033 and by the European Union NextGenerationEU/PRTR. It has also been supported by grants PID2020-112545RB-C54, PDC2022-133900-I00 and PDC2023-145862-I00, funded by MCIN/AEI/10.13039/501100011033 and by the European Union NextGenerationEU/PRTR. It has also been supported by the Spanish Ministry of Science and Innovation under Ramon y Cajal Fellowship number RYC-2020-030676-I funded by MCIN/AEI/10.13039/501100011033 and by the European Social Fund "Investing in your future" and in part by the Ministerio de Universidades, Gobierno de España under Predoctoral Grant FPU19/01251.}

\thanks{Alejandro Ramírez-Arroyo, Pablo Padilla and Juan F. Valenzuela-Valdés are with the Research Centre for Information and Communication Technologies, Department of Signal Theory, Telematics and Communications, Universidad de Granada (UGR), 18071 Granada, Spain (e-mail: alera@ugr.es, pablopadilla@ugr.es, juanvalenzuela@ugr.es).}

\thanks{Melisa López, Troels B. Sørensen and Preben Mogensen are with the Department of Electronic Systems, Aalborg University (AAU), 9220 Aalborg, Denmark (e-mail: mll@es.aau.dk, tbs@es.aau.dk, pm@es.aau.dk).}

\thanks{Ignacio Rodríguez is with the Department of Electrical Engineering, University of Oviedo (UNIOVI), 33203 Gijon, Spain (e-mail: irl@uniovi.es).}

\thanks{Samantha Caporal del Barrio and Preben Mogensen are with Nokia, 9220 Aalborg, Denmark (e-mail: samantha.caporal\_del\_barrio@nokia.com, preben.mogensen@nokia.com).}}

\markboth{Ramírez-Arroyo \MakeLowercase{\textit{et al.}}: FR2 5G Networks for Industrial Scenarios}%
{Ramírez-Arroyo \MakeLowercase{\textit{et al.}}: FR2 5G Networks for Industrial Scenarios}


\maketitle

\begin{abstract}
Industrial environments constitute a challenge in terms of radio propagation due to the presence of machinery and the mobility of the different agents, especially at mmWave bands. This paper presents an experimental evaluation of a FR2 5G network deployed in an operational factory scenario at 26~GHz. The experimental characterization, performed with autonomous mobile robots that self-navigate the industrial lab, leads to the analysis of the received power along the factory and the evaluation of reference path gain models. The proposed assessment deeply analyzes the physical layer of the communication network under operational conditions. Thus, two different network configurations are assessed by measuring the power received in the entire factory, providing a comparison between deployments. Additionally, beam management procedures, such as beam recovery, beam sweeping or beam switching, are analyzed since they are crucial in environments where mobile agents are involved. They aim for a zero interruption approach based on reliable communications. The results analysis shows that beam recovery procedures can perform a beam switching to an alternative serving beam with power losses of less than 1.6 dB on average. Beam sweeping analysis demonstrates the prevalence of the direct component in Line-of-Sight conditions despite the strong scattering component and large-scale fading in the environment.
\end{abstract}


\vspace{15mm}

\section{Introduction}
\IEEEPARstart{C}{ommunications} in factory environments are becoming of great interest in the scientific community during recent years \cite{magazine_18,magazine_19,survey_19}. Industry 4.0 and  Industrial Internet-of-Things (IIoT) aim at ensuring reliable communications under Machine-to-Machine (M2M) based scenarios \cite{iiot_1,iiot_2,iiot_3,iiot_4}. In parallel, 5G technology considers use cases based on ultra-reliable and low-latency communications (URLLC), which target end-to-end latencies below 1 ms \cite{URLLC_1,URLLC_2}. The combination of the above concepts is impacting the wireless deployments in factories, automating production and monitoring processes, while facilitating a simplification of the current industrial network layouts \cite{automation_1,automation_2}.

To provide efficient and reliable communications in factory environments, it is fundamental to deeply analyze radio propagation phenomena in these environments. Millimeter waves (mmWaves)/Frequency Range 2 (FR2) arise as a suitable solution due to the higher bandwidth available, which implies higher channel capacity, less interference between frequencies, and higher localization accuracy due to the shorter wavelength compared to sub-6 GHz/Frequency Range 1 (FR1) \cite{mmWave}. However, industrial scenarios prove to be challenging environments since radio propagation at mmWaves may exhibit disruptive characteristics for communications due to:

\begin{enumerate}[label=(\roman*)]
  \setcounter{enumi}{0}

  \item High attenuation and blockage probability: mmWaves undergo high attenuation. This effect is enhanced by the presence of metallic structures, machinery and equipment that generate a propagation channel rich in multipath components \cite{blockage_factories}.
  
  \item Variations in the environment and mobility from the agents involved: the scenario is time-dependent due to the variable conditions because the mobile agents involved are constantly changing position \cite{mobility_agents}.
  
  \item Network deployment design and configuration: antenna gain, directivity and radiation pattern are critical parameters. Optimal design and configuration can significantly improve the coverage for mmWave communications \cite{Valenzuela}.
  
\end{enumerate}

\begin{figure*}[t]
    \centering
    \subfigure[]{\includegraphics[width= 0.322\textwidth]{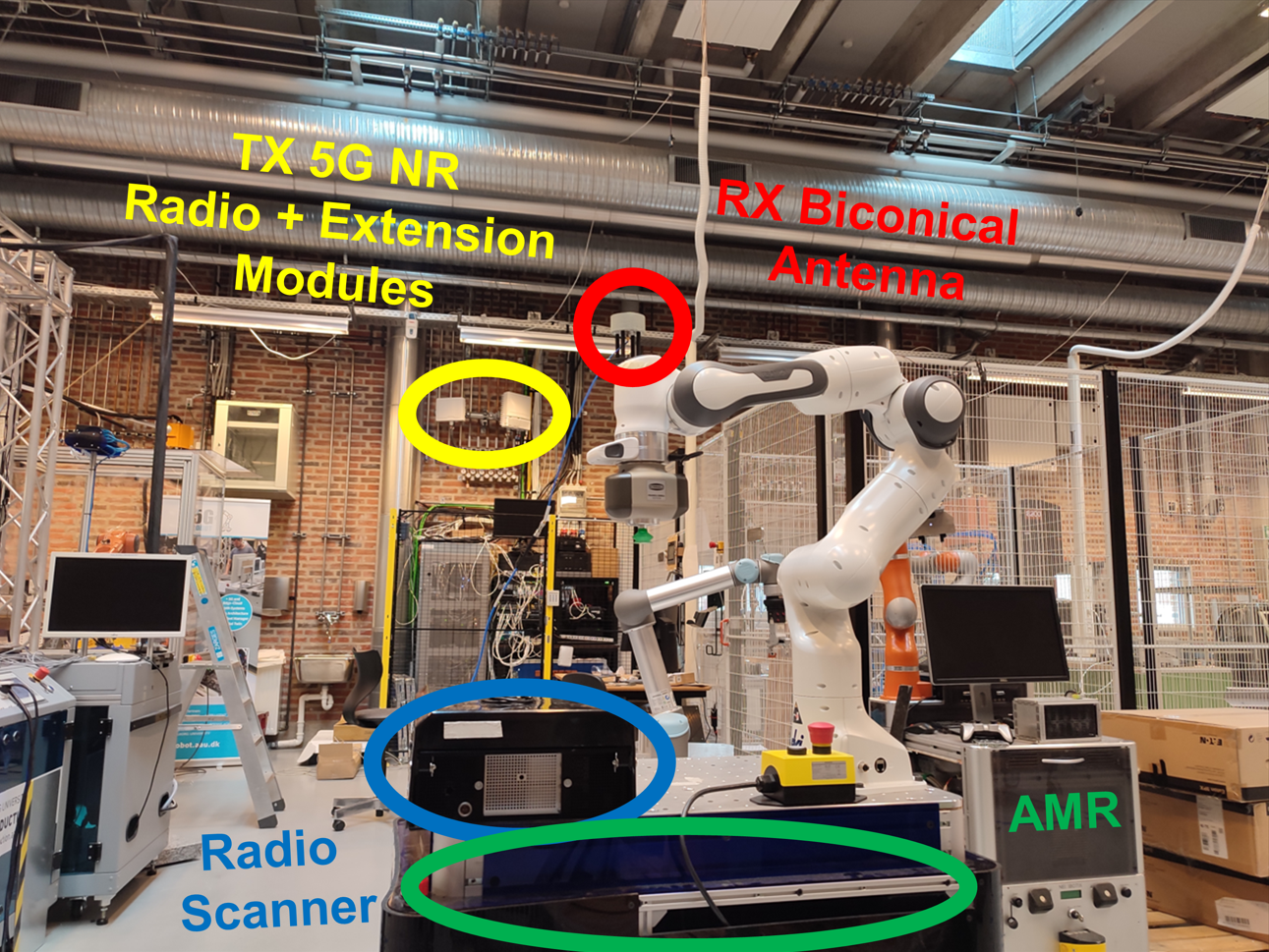}
	}
	\subfigure[]{\includegraphics[width= 0.322\textwidth]{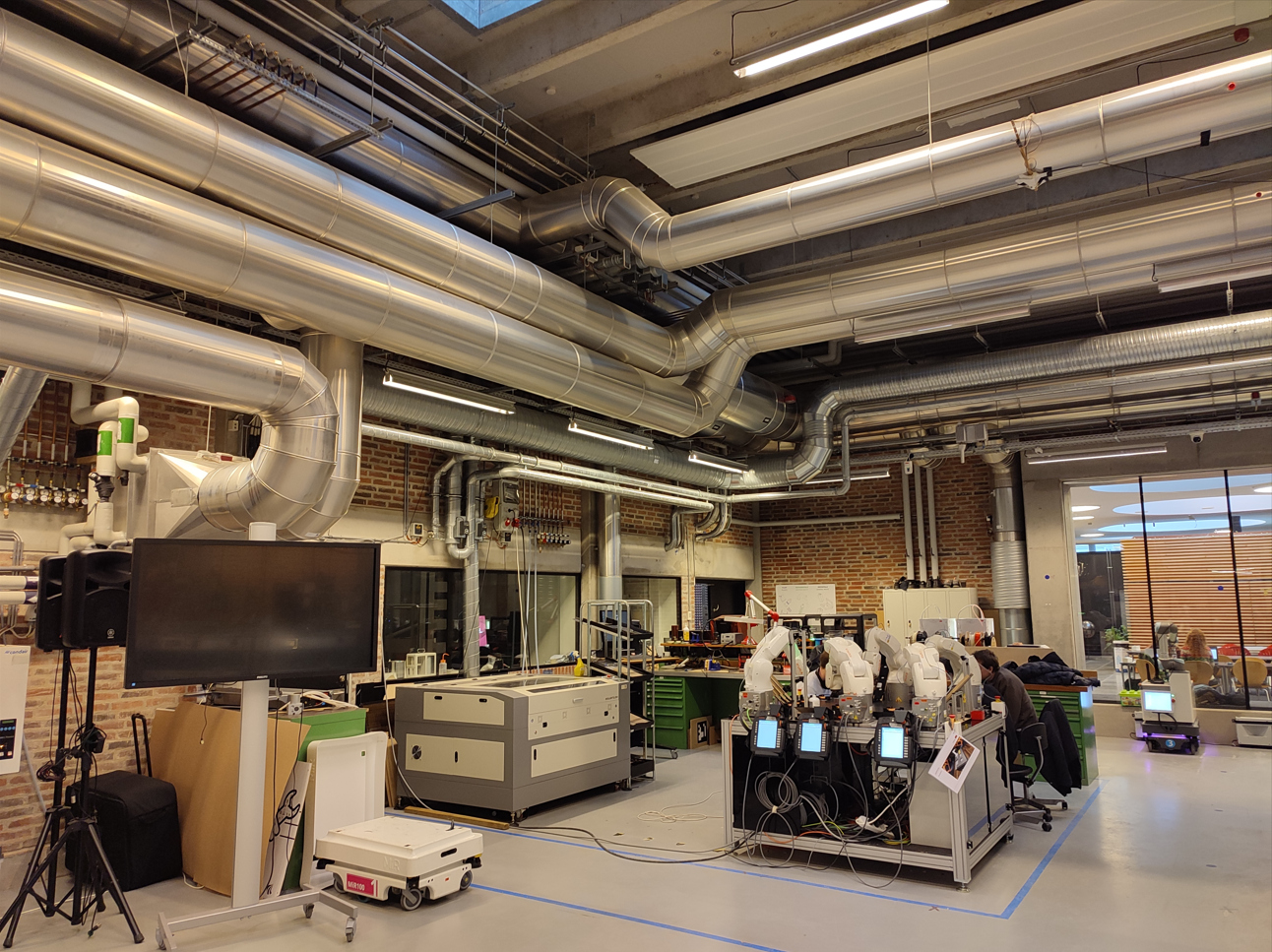}
	}
	\subfigure[]{\includegraphics[width= 0.322\textwidth]{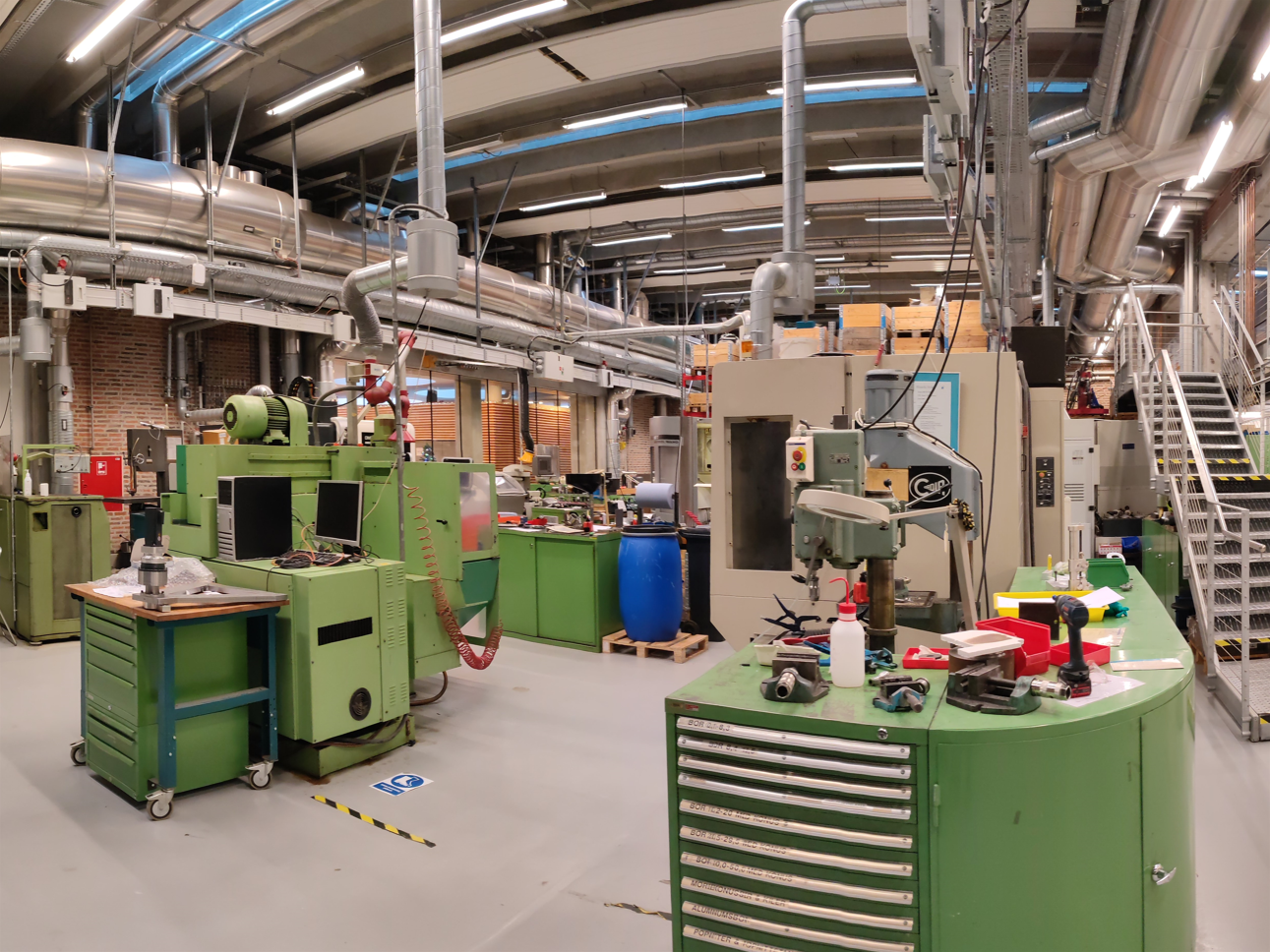}
	}
    \caption{Photographs of (a) the measurement equipment at both sides, TX (wall-mounted FR2 micro BS) and RX (AMR-mounted radio scanner), (b) the view of the industrial automation and robotics clutter in the sparse hall, and (c) the view of the heavy machinery in the dense hall.} 
	\label{fotos}
\end{figure*}

The previous issues can be solved by performing analog beamforming and smart allocation of radio resources \cite{beamforming, resource_allocation}. Highly directive beams can compensate for high attenuation losses. Combined with analog beamforming, mobile agents can be tracked to maximize coverage. In 5G NR, all these procedures are included in the beam management architecture, such as, recovery, sweeping, measurement, or switching \cite{5G_physical_layer}. In order to successfully apply the described technical solutions in industrial scenarios, detailed knowledge about the varying radio propagation behavior of the different beams is needed, especially to guarantee service continuity when addressing demanding services in terms of high throughput and reliability under operational mobility conditions. Summarizing the state-of-the-art regarding industrial wireless communications, Khoshnevisan \textit{et al.} \cite{automation_1} propose a network architecture for industrial factory automation given a coordinated multipoint scheme in the context of ultra-reliability
and low-latency communications.  Ghatak \textit{et al.} \cite{blockage_factories} present a stochastic framework channel model given the high probability of physical blockages in the industrial environment. Fortino \textit{et al.} \cite{mobility_agents} show a framework for coordinating several mobile agents in factory-based scenarios. Chizhik \textit{et al.} \cite{Valenzuela} present measurement campaigns based on directional measurements to characterize the propagation channel within factory scenarios. Finally, it is noteworthy that organizations such as  3GPP are starting to standardize channel models for indoor factory environments \cite{3GPP_CI, tr38901}. While most of the works found in the literature has a theoretical approach \cite{automation_1, blockage_factories, mobility_agents}, the experimental works \cite{Valenzuela, 3GPP_CI} consider a specific and dedicated setup for the analysis. As it will be presented, one of the main novelties of this work is the analysis based on the operational conditions of the industrial environment, where autonomous mobile robots are self-driving throughout the factory-based scenario given a 5G NR network~deployment.

In this work, an experimental evaluation of a 5G operational network at FR2 frequencies in a factory environment is performed. This study illustrates the previously described issues and analyzes FR2 coverage in a small operational factory, considering two different network deployment configurations. The study surveys different coverage-related aspects leading to the optimal BS configuration and cell design, and benchmarks the experienced radio propagation levels with those from reference path loss models typically applied to industrial scenarios, evaluating their suitability in this type of environments. Further, the study considers beam-specific analysis to assess the quality of service given a specific serving beam, and study beam management procedures relevant for mobile elements such as (i) beam sweeping and switching, which is fundamental for analyzing the spatial distribution of the beams in the factory; (ii) and beam recovery, which allows signal recovery in the event that a specific beam is affected by high attenuation.

In particular, the novelties of the study are:

\begin{itemize}

\item Comparison of two measurement campaigns carried out at the same factory for two different base station configurations under operational conditions.

\item Development of design rules from signal strength coverage maps based on the two operating conditions: These coverage maps allow estimating a cell size for deployments in industrial environments, as well as evaluating the path gain based on state-of-the-art reference models.

\item Analysis of beam management procedures: In particular, a study of beam recovery, beam sweeping, and beam switching procedures is carried out based on the measurements performed with autonomous mobile robots that self-navigate the factory.

\item Beam switch-off optimization study: Based on the coverage maps generated from the self-navigation agents, it is proposed to turn off a subset of beams, looking for a trade-off between the coverage areas and the number of beams turned on simultaneously. Therefore, this study may lead to the beam switch-off as an energy-saving solution.

\end{itemize}

The results are expected to be useful for academic researchers and engineers for radio planning of similar industrial hall scenarios based on FR2 wall-mounted directional antennas, taking into account not only propagation aspects but also mobility and beam management.

The work is organized as follows. Section II presents the industrial lab test environment, the measurement equipment, and the measurement campaigns performed. Section III analyzes the radio propagation aspects in the industrial scenario. Section IV studies the beam management procedures involved in industrial communications. Section V shows the coverage maps according to the beam management procedures and a beam switch-off study. Finally, Section VI summarizes the main contributions of this document.

\section{Industrial Scenario and Acquisition Process for 5G NR}

\subsection{Measurement scenario}

The present study is conducted at the experimental setup formed by the 5G Smart Production Lab in Aalborg University (AAU), Denmark \cite{magazine_smartlab}. This industrial research lab resembles a small factory with an extension of approximately 1350 $\textrm{m}^{2}$, equipped with real manufacturing and production equipment. The factory environment is formed by two different halls, hereinafter referred to as sparse and dense halls. The sparse hall, with dimensions $\textrm{15 m} \times \textrm{40 m} \,\, (\textrm{600 m}^2)$ and shown in Figs.~\ref{fotos}(a) and \ref{fotos}(b), contains wide corridors and metal machinery with heights ranging from $1.5$ m to $2.5$ m. The dense hall, with dimensions $\textrm{25 m} \times \textrm{30 m} \,\, (\textrm{750 m}^2)$ and shown in Fig. \ref{fotos}(c), consists of a larger number of heavy machinery with heights between $2.5$ m and $3.5$ m. The higher density of machinery results in narrower aisles. In both cases, the radio propagation conditions are determined by a strong multipath effect due to reflections and scattering in the heavy machinery~\cite{Alex_EuCAP}.

\begin{figure}[t]
\centering
\includegraphics[width= 1\columnwidth]{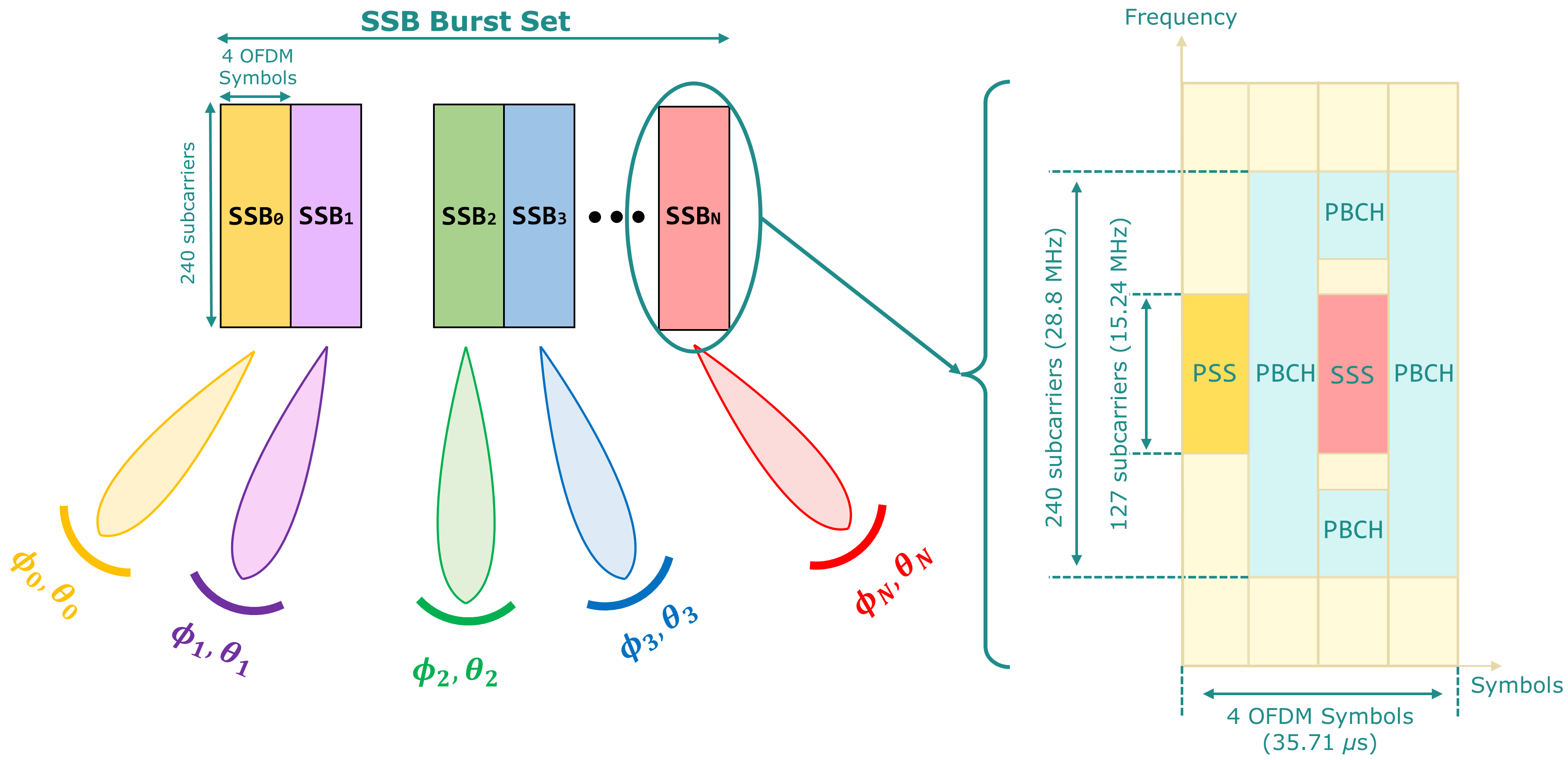}
\caption{Synchronization Signal Block structure and beamforming through the burst set.}
\label{SSB_block}
\end{figure}

\subsection{5G NR Physical channel and measurement equipment}

In 5G NR, physical time-frequency resources are organized in orthogonal frequency-division multiplexing (OFDM) symbols and subcarriers for the time and frequency domains, respectively. In the frequency domain, a physical resource block (PRB) is formed by 12 subcarriers, while 14 consecutive OFDM symbols form a slot in the time domain \cite{ts38211}. In order to provide a wide range of configurations, 5G NR allows OFDM subcarrier spacing (SCS) of $15 \times 2^n$ kHz where $n$ is an integer value. This flexible numerology determines the slot duration, given by $1/2^n$ ms, and consequently the OFDM symbol duration. In particular, 3GPP TS 38.211 and 3GPP TS 38.213 define up to five use cases from A to E according to the subcarrier spacing and the frequency range \cite{ts38211, ts38213}. In this work, the analysis is performed for case D, which is expected to be employed in 5G NR taking into account a subcarrier spacing of $120$~kHz $(n= 3)$ and operating frequencies in the FR2-1 range ($24.25$~GHz - $52.6$~GHz). Thus, OFDM symbol duration with cyclix prefix is fixed to $8.91\,\mu$s.

In the physical layer, the synchronization signal (SS) and physical broadcast channel (PBCH) are enclosed in the same block, known as synchronization signal block (SSB). This block, depicted in Fig.~\ref{SSB_block}, is formed by 4 OFDM symbols and 240 subcarriers. Specifically, primary (PSS) and secondary (SSS) synchronization signals expand over 127 subcarriers. For high frequencies, where narrow beams can be considered, SSBs are beamformed in order to provide good spatial coverage, as it is shown in Fig.~\ref{SSB_block}. Therefore, these blocks are beamformed and transmitted periodically and repeatedly through SSBs burst sets. The periodicity and the burst set duration are defined in the 3GPP TS 38.101-4 standard \cite{ts381014}. In this work, these values are configured to $20$~ms and $5$~ms, respectively. The periodicity of the SSBs can be used for beam management tasks, as will be shown in later sections.

\begin{figure}[t]
    \centering
    \subfigure[]{\includegraphics[width= 0.48\columnwidth]{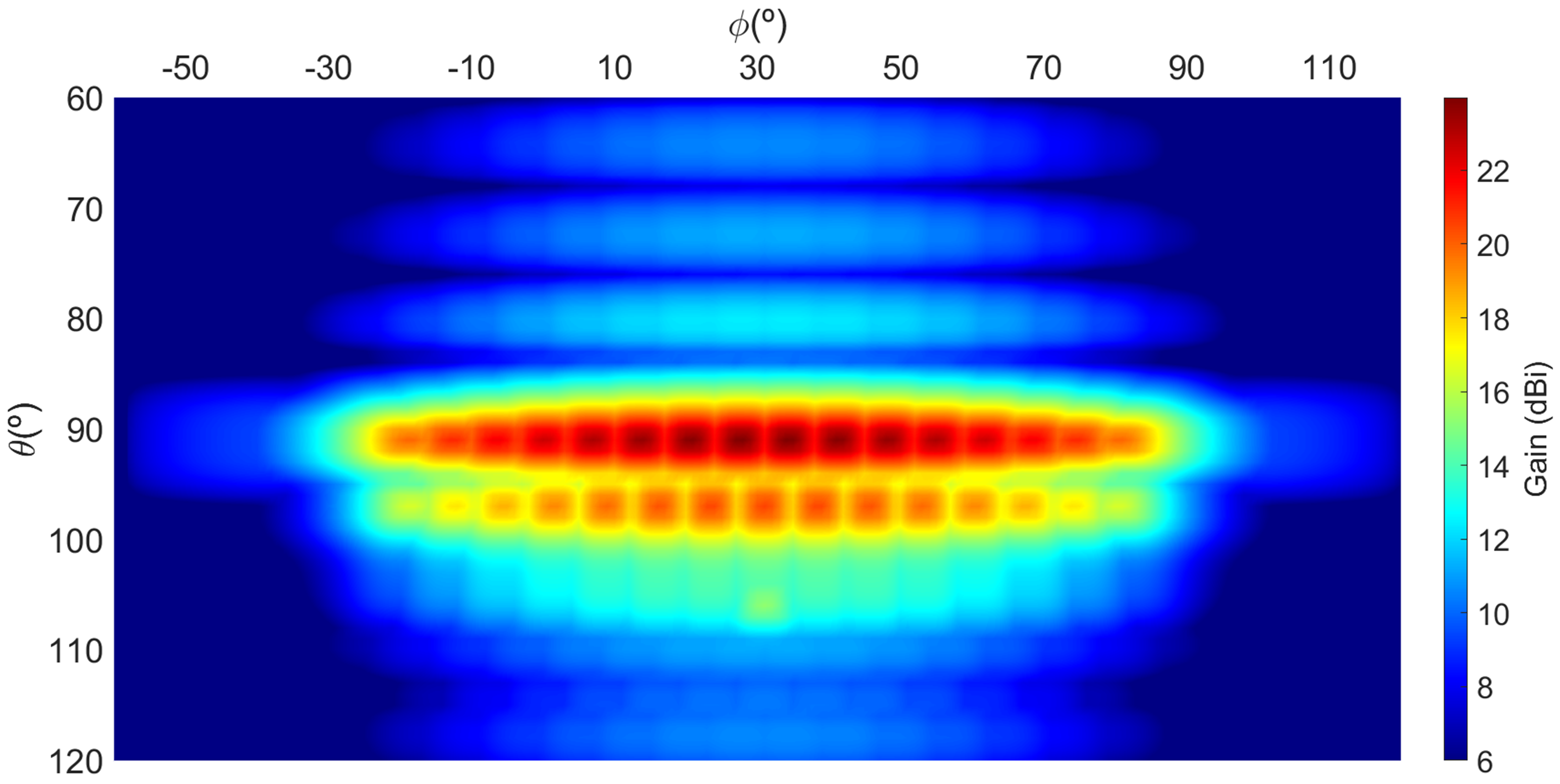}
	}
	\subfigure[]{\includegraphics[width= 0.48\columnwidth]{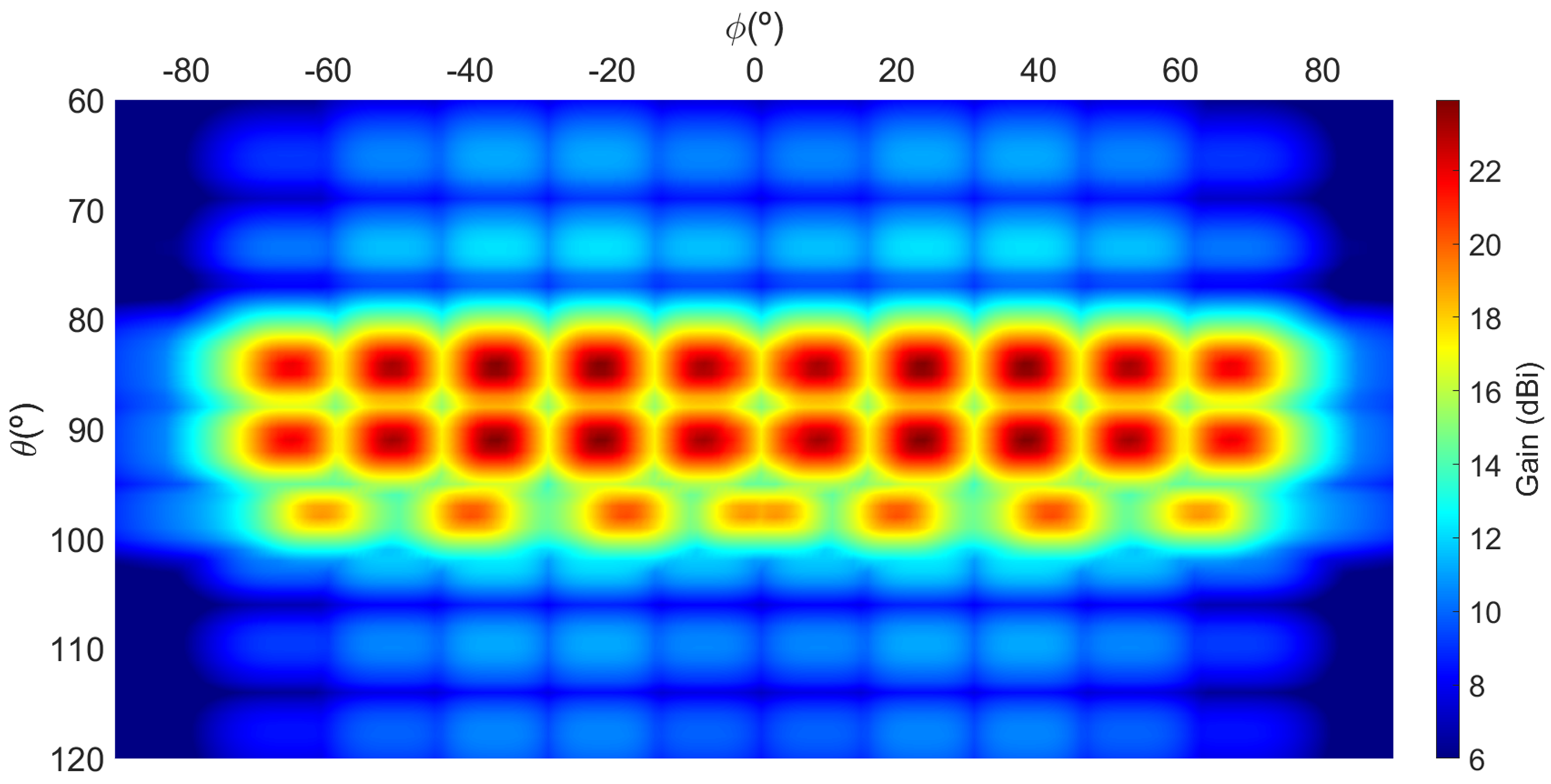}
	}
    \caption{SSB beamforming for the TX module in (a) configuration A, and (b)~configuration B.} 
    \label{patterns}
\end{figure}

\renewcommand{\arraystretch}{1.5}
\setlength{\tabcolsep}{4pt}

    \begin{table}
    \centering
    \caption {TX Module Parameters in terms of Configuration} \label{tabla_1} 
    \resizebox{\columnwidth}{!}{

    \begin{tabular}{c|c|c}
    \hline \hline
    \multicolumn{1}{c|}{Parameter} & \multicolumn{1}{c|}{Configuration A} & \multicolumn{1}{c}{Configuration B}\\
    \hline \hline
     
     
    \hline Number of beams &  32 & 27\\
    \hline Beams per elevation angle &  16,15 and 1 & 10, 10 and 7\\
    \hline $\phi$ and $\theta$ aperture & $120\degree$ and $15\degree$  & $150\degree$ and $15\degree$\\
    \hline Coverage scanning & \makecell{$\phi \in [-30\degree, 90\degree]$ \\ $\theta \in [90\degree, 105\degree]$} & \makecell{$\phi \in [-75\degree, 75\degree]$ \\ $\theta \in [83\degree, 98\degree]$}\\
    \hline \hline

    \end{tabular}

    }
    
\end{table}

In order to statistically characterize the factory propagation environment, an R\&S TSME6 radio scanner \cite{TSME6} is employed to capture radio parameters. In particular, the radio scanner measures the Reference Signal Received Power (RSRP) as the average power over the SSS bandwidth ($15.24$ MHz for SCS~$= 120$ kHz) given a specific SSB. This scanner, shown in Fig.~\ref{fotos}(a), is mounted on the top of an Autonomous Mobile Robot (AMR), which can be configured to follow routes throughout the factory. Additionally, the AMRs are configured to self-navigate the industrial lab, following predefined routes and missions. This fact, initially designed to maximize the performance in industrial processes involving mobile robots~\cite{AMR_industrial}, simplifies the radio propagation analysis by automating the measurement process, ensuring route repeatability with cm-level accuracy. Thus, the RSRP can be acquired given the different locations of the AMR in the factory. In order to capture the SSB from the network, the TSME6 is configured with a biconical receiver (RX) antenna \cite{RX_antenna} mounted on the top of the robotic arm, illustrated in Fig.~\ref{fotos}(a), with ultra wide-band bandwidth ranging from $3$~GHz to $40$~GHz. The RX height is $1.5$ m above the floor. Note that the maximum AMR speed is 2 m/s, which implies a maximum frequency shift of 173 Hz considering the operating frequencies. This shift is negligible according to the frequency carrier in the FR2 band and the subcarrier spacing \cite{Doppler}.

The 5G NR network under test consists of a transmitter (TX) based on a Nokia 5G Airscale mmWave Radio module~\cite{TX_module}. This module operates in the FR2 n258 band ($24.25$~GHz - $27.5$~GHz) according to case D previously described in this subsection. By using an extender module, that can be used in combination with the main 5G radio module, a total of 32 beams with different azimuth and elevation angles can be generated. The horizontal beamwidth ranges from $8\degree$ to $12\degree$, while the vertical beamwidth goes from $6\degree$ to $10\degree$, thus providing high directivity beams covering several spatial regions in the factory.

\begin{figure*}[!t]
    \centering
    \subfigure[]{\includegraphics[width= 0.324\textwidth]{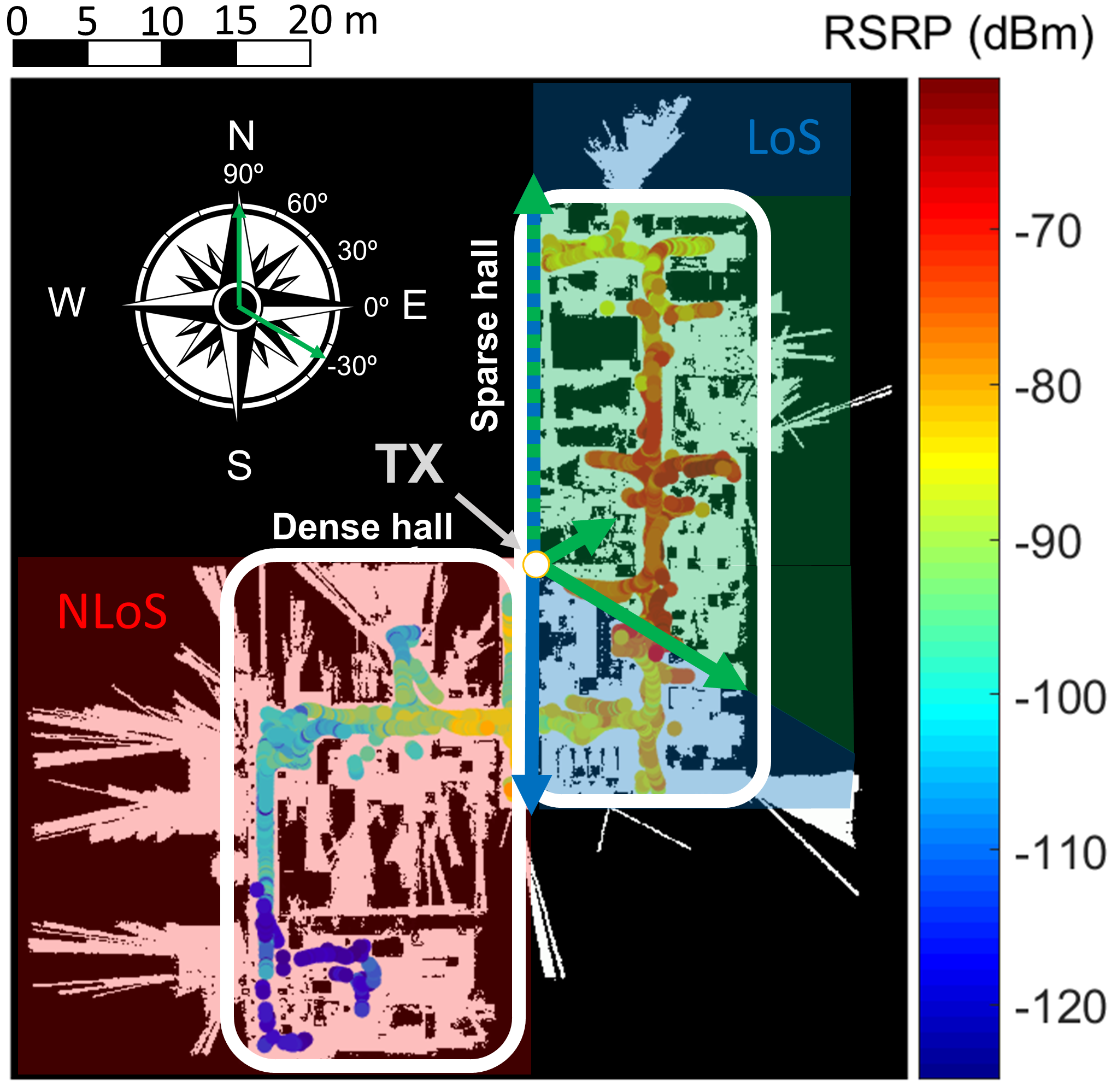}
	}
	\subfigure[]{\includegraphics[width= 0.324\textwidth]{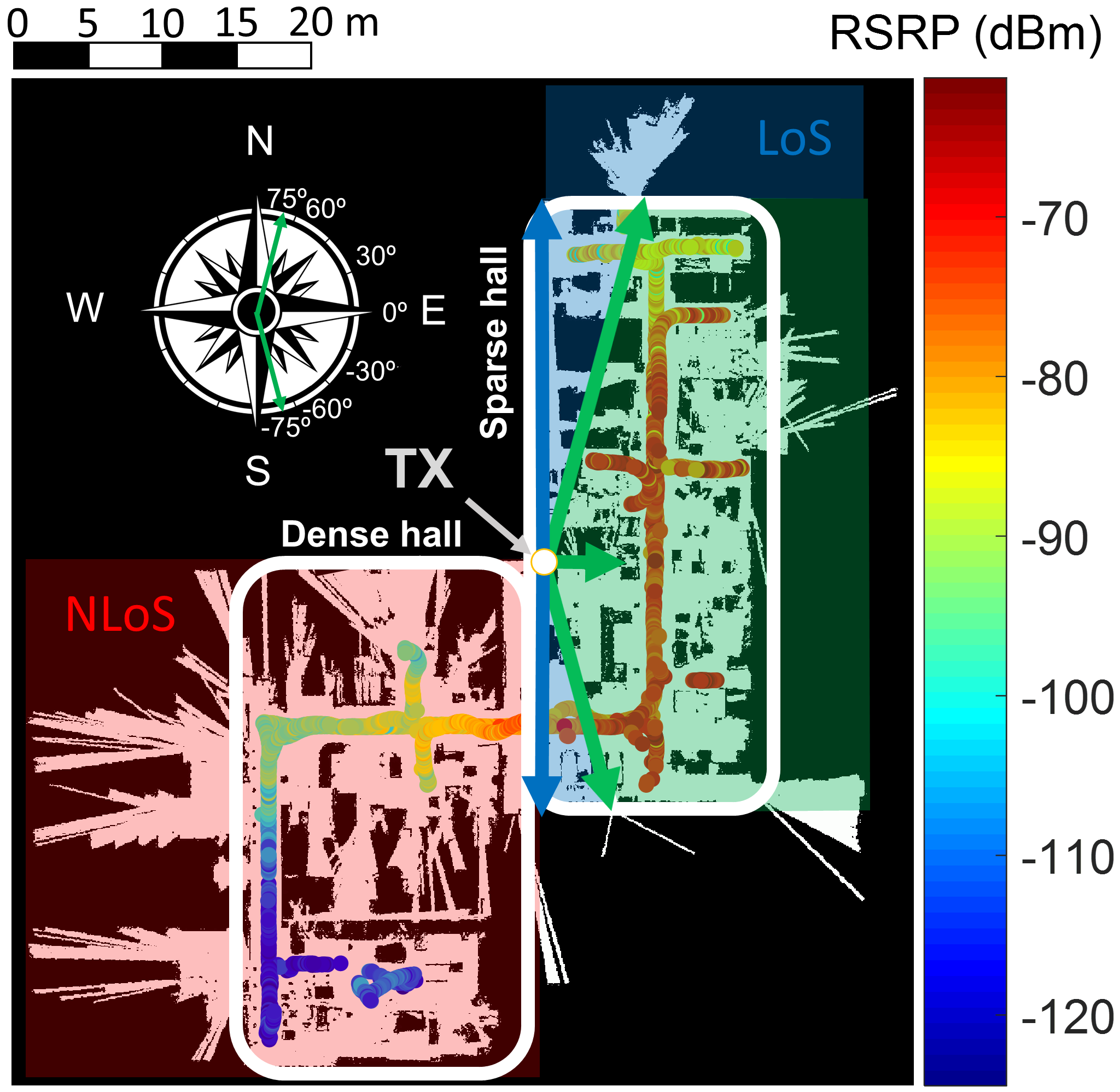}
	}
	\subfigure[]{\includegraphics[width= 0.317\textwidth]{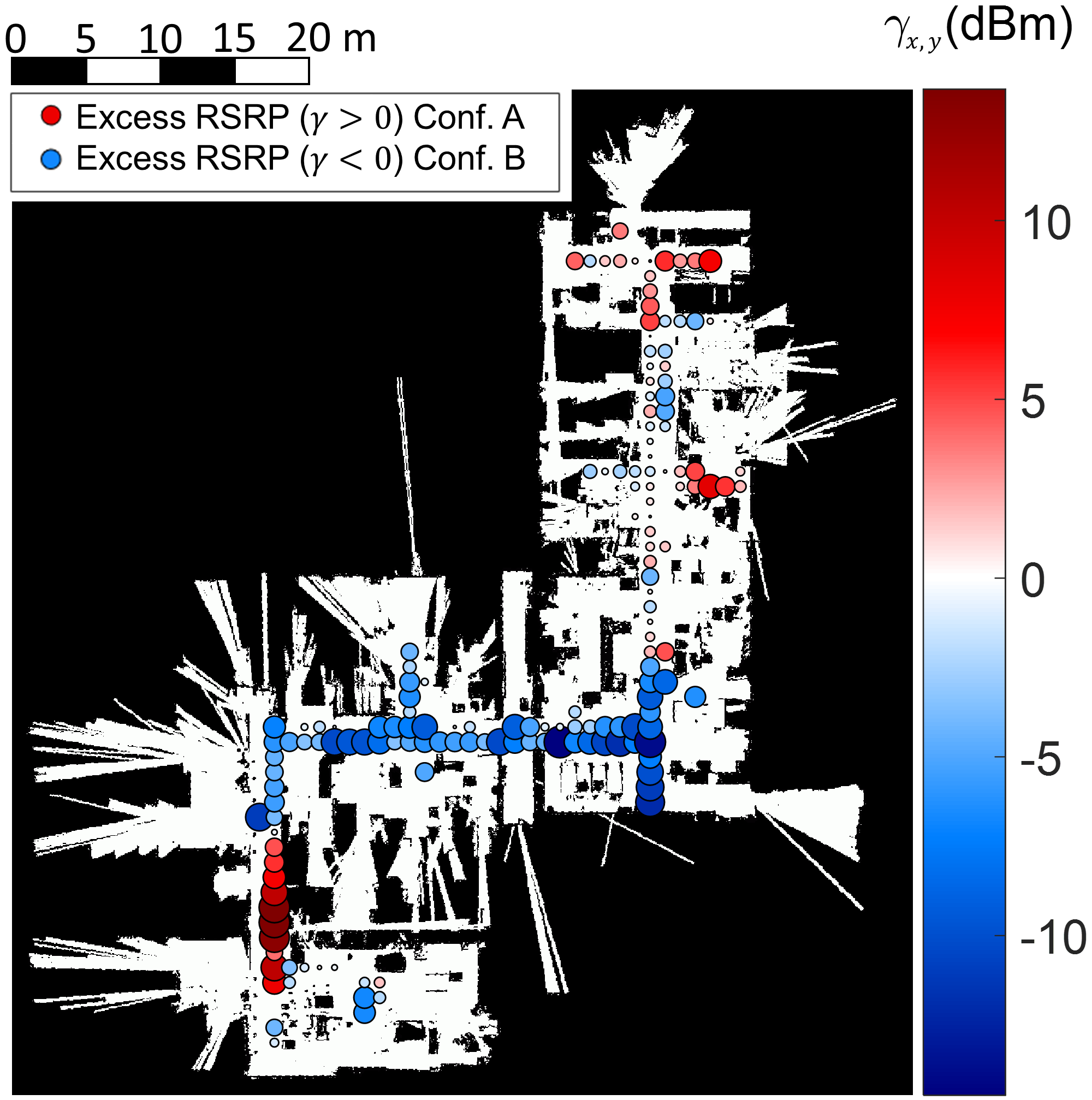}
	}
    \caption{Maximum SS-RSRP received at each position in the factory for (a) configuration A and (b) configuration B. (c) Difference of the average SS-RSRP in favor of each of the configurations. There is LoS condition in the sparse hall (blue area) and NLoS condition in the dense hall (red area). Areas under antenna boresight influence are marked in green.} 
	\label{RSRP_map}
\end{figure*}

In this work, two TX module configurations are analyzed in depth:

\begin{itemize}

    \item Configuration A (illustrated in Fig. \ref{patterns}(a)): The main module with 32 beams covering a horizontal angle of $120\degree$ is considered. These are distributed in three elevation angles with $0\degree$, $7\degree$ and $15\degree$ downtilt, respectively.
    
    \item Configuration B (shown in Fig. \ref{patterns}(b)): The main module operates together with the extension module, providing a total of 27 effective beams with $150 \degree$ horizontal scanning. In this case, beams are distributed in three elevation angles. Specifically, $-7\degree, 0\degree$ and $8\degree$ downtilt values are considered.
    
\end{itemize}

  Table~I summarizes the main parameters for each configuration. Note that $\phi$ and $\theta$ stand for the horizontal and vertical angles, respectively. Configuration A has a higher density of beams in the horizontal scan. This causes the angular distance between adjacent beams to be minimal. Configuration B, on the other hand, has a greater separation of beams, allowing it to cover almost the entire sparse hall. The effects of both configurations will be discussed in later sections. For the sake of clarity, throughout the work, beams are numbered according to their corresponding $\textrm{SSB}_{\zeta,r,c}$, where $\zeta$ stands for the module configuration, and $r$ and $c$ are the beam row (from top to bottom) and column (from left to right), respectively.

\section{Radio Propagation Analysis}

This Section presents several aspects related to the FR2 5G Network deployed in the 5G Smart Production Lab under operational conditions. A comparison is made between the two configurations of the TX module presented in Section II.B. The SS-RSRP analysis along the industrial environment allows to estimate the coverage and define a maximum cell radius according to a threshold target. Both measurement campaigns are used to fit path gain models in Line-of-Sight (LoS) and Non Line-of-Sight (NLoS) condition, which are validated with state-of-the-art channel models for indoor industrial scenarios. The correct modeling of the previous aspects is fundamental for the radio planning of wireless communications systems.

\subsection{RSRP analysis}

To analyze the nature of the propagation environment, the AMR is configured to cover several routes across the factory with $1.5$ m/s average speed. Every 20 ms, the radio scanner detects the SS-RSRP coming from each SSB in the burst set. Figs. \ref{RSRP_map}(a) and \ref{RSRP_map}(b) show the strongest SS-RSRP for each burst set on each TX configuration. Note that the TX module is located on the west side of the sparse hall at a height of $3$ m. Given this location, the factory is divided into two areas depending on the visual line of sight between TX and RX: Line-of-Sight and Non Line-of-Sight. In Fig.~\ref{RSRP_map}(a), the area under the boresight influence experiences SS-RSRP values between $-62$~and $-80$~dBm. In the southern part of the sparse hall, the decrease of the SS-RSRP is remarkable when leaving the influence of the beams. Finally, in the east-west-oriented corridor the SS-RSRP ranges between $-85$~dBm and $-100$~dBm, while in the southern part of the dense hall, completely obstructed by the heavy machinery, the SS-RSRP is below $-100$~dBm.

For the configuration B, depicted in Fig. \ref{RSRP_map}(b), the boresight of the beams almost completely covers the sparse hall. This causes the southern part of the sparse hall to be covered with SS-RSRP values up to $-75$ dBm. Additionally, due to reflections and scattering from metallic structures such as pipelines and ventilation ducts, shown in Fig. \ref{fotos}(b), the east-west-oriented corridor SS-RSRPs are in the range of $-75$~dBm to $-95$~dBm. 

\begin{figure}[t]
    \centering
    \subfigure[]{\includegraphics[width= 0.99\columnwidth]{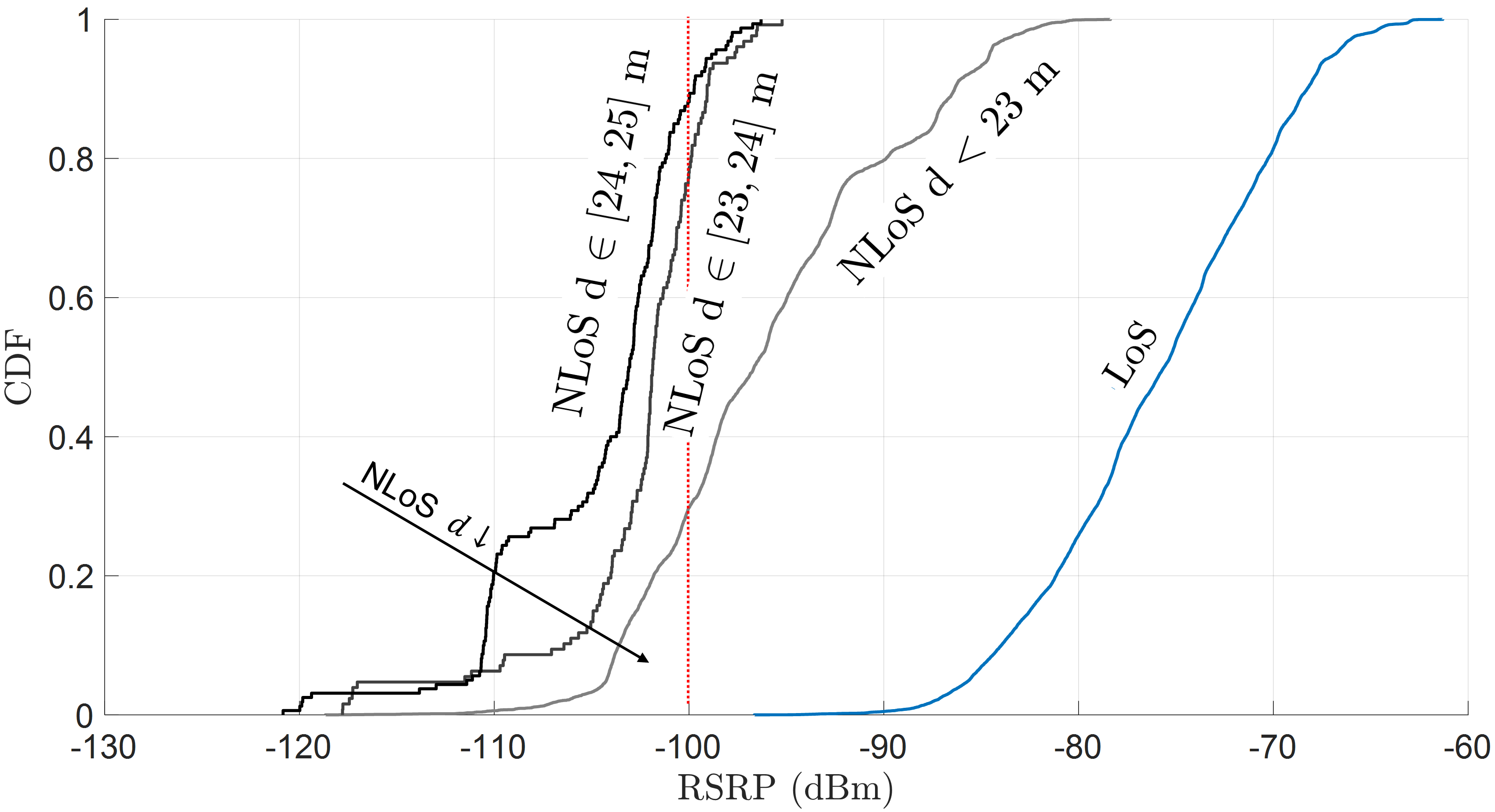}
	}
	\subfigure[]{\includegraphics[width= 0.99\columnwidth]{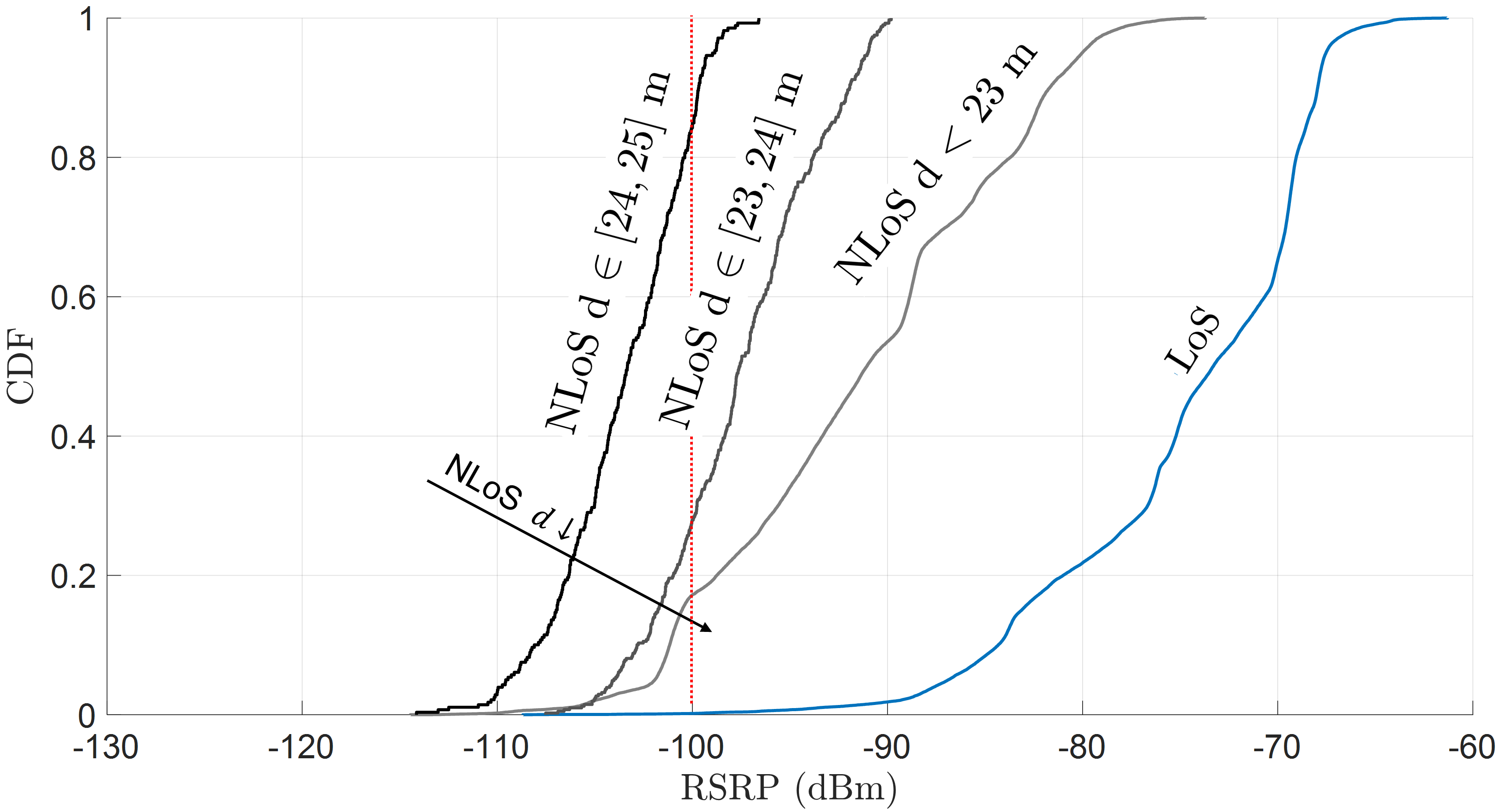}
	}
\caption{RSRP CDF for LoS and NLoS regions in (a) configuration A and (b) configuration B. NLoS regions are divided into several distance ranges.}
\label{RSRP_cell_coverage}
\end{figure}

In order to compare both configurations, the factory is divided into a squared grid of size $1 \times 1$~$\textrm{m}^2$. The SS-RSRP of all points belonging to each square is averaged. In terms of wavelengths, the averaging is performed on squares of dimensions $86.7\lambda \times 86.7\lambda$, which makes it possible to obtain the local average SS-RSRP while removing the fast fading effect \cite{Lee_criteria}. The comparison between both values of each campaign indicates whether the radio propagation is more favorable in one case or the second one. Fig. \ref{RSRP_map}(c) shows the difference between averaged SS-RSRP values on each square in the grid. Therefore, $\gamma_{x,y}$ is defined as:

\begin{equation}
\gamma_{x, y}=\overline{RSRP}_{A, x, y}-\overline{RSRP}_{B, x, y},
\end{equation}

\noindent where $\overline{RSRP}_{A, x, y}$ and $\overline{RSRP}_{B, x, y}$ stand for the averaged SS-RSRP in the coordinates $(x,y)$ for configuration A and B, respectively. Thus, $\gamma_{x, y}$ indicates the difference in terms of SS-RSRP for both configurations at each location. Red colors mark areas where the SS-RSRP is stronger in configuration A compared to configuration B. The blue colors indicate areas where coverage is better in configuration B than in configuration A. 
 
 It can be concluded that, even when the beams in configuration B do not impinge directly on the corridor connecting both halls, there is a noticeable improvement in the SS-RSRP due to the better coverage of the southern part of the sparse hall. This is a direct consequence of the higher separation of the beams in the azimuthal direction. This separation does not negatively affect the range $\phi \in [-30\degree,75\degree]$ originally covered by configuration A, since the values of $\gamma_{x,y}$ in this range are close to zero. Thus, considering the average spatial coverage given by the SS-RSRP from the strongest beam, configuration~B is considered a better solution for network deployment.

\subsection{Cell coverage}

Regarding network deployment, the cell size is fundamental in order to determine the number of access points required to provide service on the downlink. The cell coverage in the network can be estimated as the service availability given a certain cell radius $r$. For a fixed cell radius $r$, if the RSRP is above a certain threshold, good quality of service can be guaranteed. Assuming a RSRP threshold of -100 dBm \cite{service_availability}, an estimation of the maximum size of the cell can be made through the probability $P(\textrm{RSRP} < -100 \textrm{ dBm} \,\,|\,\,r)$. For that purpose, Fig.~\ref{RSRP_cell_coverage} shows the RSRP CDF for several distances and both configurations. In Figs.~\ref{RSRP_cell_coverage}(a) and (b), the RSRP is higher than -100 dBm in 99.8\% of the measurements for any distance in the LoS region. Given that the maximum distance measured in the LoS case is 26 meters, a similar cell radius could be considered with the assurance of providing good coverage. In the NLoS region, 29.8\% and 17.3\% of measurements are below the threshold when distances lower than 23 meters are considered. These values increase drastically when considering measurements in the ranges $d \in[23, 24]$ m and $d \in [24, 25]$ m, with percentages of 77.9\%, and 88.1\% (configuration A) and 27.9\%, and 84.9\% (configuration B). Therefore, cell radii of less than 23 meters should be used in the design of 5G NR networks at mmWave frequencies in industrial environments with NLoS assuming a target of $\textrm{RSRP} > -100$~dBm. Note that the RSRP improvement illustrated in Fig. \ref{RSRP_map} for configuration~B results in a shift of the CDFs to the right, which implies lower probabilities for $P(\textrm{RSRP} <~-100 \textrm{ dBm} \,\,|\,\,r)$. From the capacity perspective, it has been demonstrated that data rates higher than 130 Mbps can be obtained in industrial environments in the FR2 range for 90\% coverage, taking into account ranges up to 64 m \cite{Valenzuela}.

\subsection{Path Gain analysis}

To estimate the Path Gain (PG) in the factory environment, it is calculated from the SS-RSRP as:

\begin{equation}
    PG\textrm{[dB]} = RSRP\textrm{[dBm]} - P_{TX}\textrm{[dBm]} - G_{TX}\textrm{[dB]} - G_{RX}\textrm{[dB]},  
\end{equation}

\noindent where $G_{TX}$ and $G_{RX}$ are the nominal gain for the TX and RX antennas given $\phi$ and $\theta$ angles for a specific TX-RX link. $P_{TX}$ is the transmitted power over a resource element since RSRP is referred to the power contribution of a single resource element. Given a total transmission power over $100$ MHz carrier bandwidth of $P_{c} = 21.2$~dBm, the transmitted power over a resource element can be calculated as $P_{c}/N_{RE}$ where $N_{RE}$ is the number of resource elements (subcarriers) in the carrier bandwidth. According to 3GPP TS 38.104 \cite{ts38104}, for $120$ kHz subcarrier spacing and $100$~MHz carrier bandwidth, $N_{RE} = 66 \cdot N_{RB} = 792$. Additionally, as a consequence of scattering effects and obstruction in the factory, the effective gain of high directivity antennas is degraded, especially in NLoS conditions. Thus, the nominal gain of TX is compensated by 1~dB in LoS and 4.9~dB in NLoS to account for the typical median effective gain degradation in factories at mmWave frequencies \cite{Valenzuela}. 

To model the scenario, a generic slope intercept model is proposed as:

\begin{equation}
P G(d)=P G_{1 m}-10 n \log _{10}(d)+\mathcal{N}\left(0, \sigma^2\right),
\end{equation}

\noindent where $PG_{1m}$ is the path gain at $1$ m distance, $n$ is the path loss exponent which determines the slope given $d$ and $\sigma$ is the standard deviation for the model, stating the goodness of fit for the model.

\begin{figure}[t]
    \centering
    \subfigure[]{\includegraphics[width= 1\columnwidth]{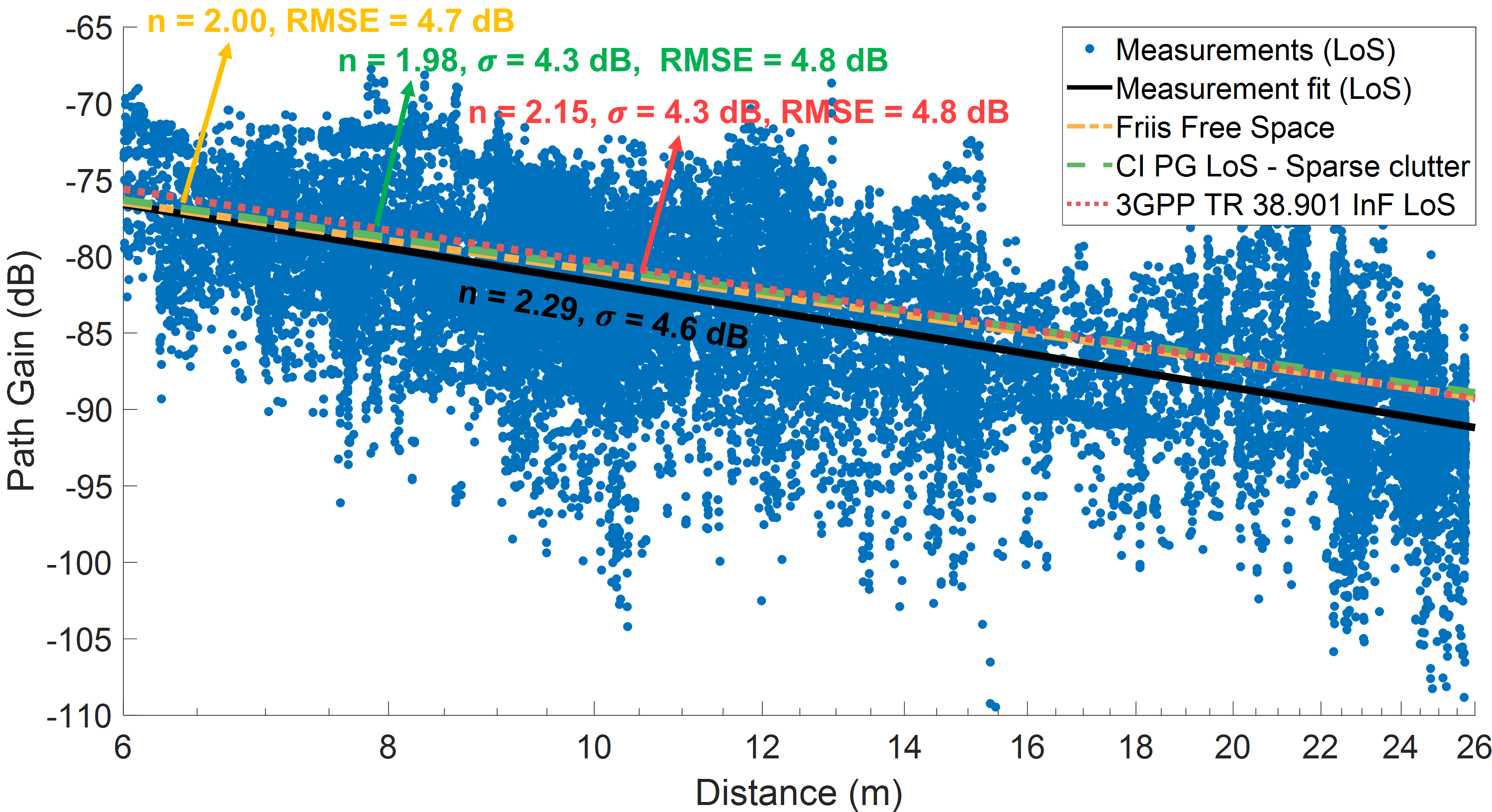}}
	
	\subfigure[]{\includegraphics[width= 1\columnwidth]{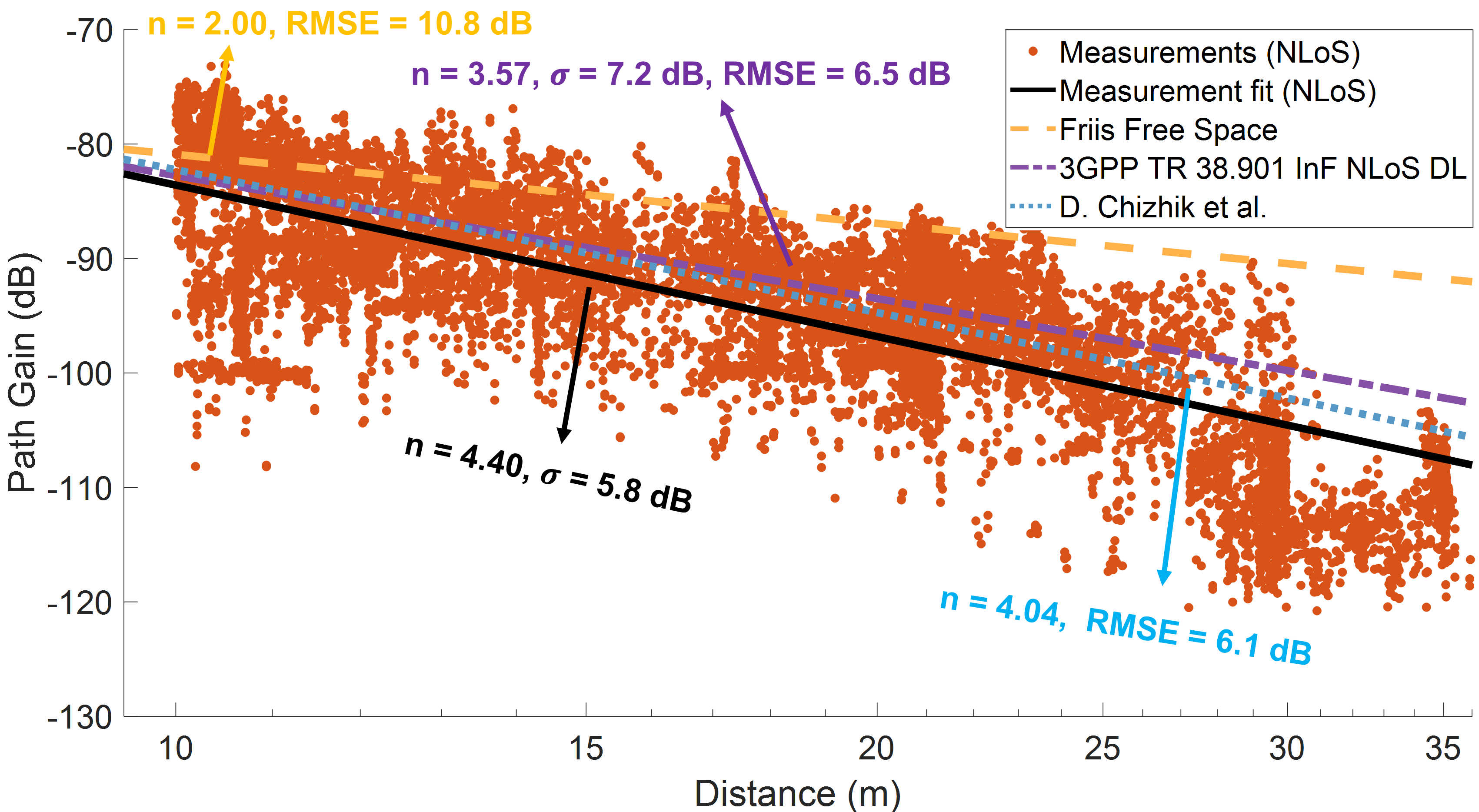}}
	
    \caption{Path Gain in terms of the distance for the measurements performed in both campaigns in: (a) LoS and (b) NLoS condition. Several slope intercept models are included for comparison purposes.} 
	\label{PG}
\end{figure}

\renewcommand{\arraystretch}{1.5}
\setlength{\tabcolsep}{4pt}

\begin{table}
    \centering
    \caption {Slope intercept Path Gain models for factories \\ in LoS and NLoS} \label{tabla_2} 
    \resizebox{\columnwidth}{!}{

    \begin{tabular}{c|c|c|c}
    \hline\hline
    \multicolumn{4}{c}{Line-of-Sight} \\
    \hline \hline
    \multicolumn{1}{c|}{PG model} & \multicolumn{1}{c|}{$PG_{1m}$} & \multicolumn{1}{c|}{$n$} & \multicolumn{1}{c}{$\sigma$ $\backslash$ RMSE}\\
    \hline \hline
     
     
    \hline Measurement Fit (LoS) &  $-58.8$ & 2.29 & $\!4.6$ $\backslash$ $\;\;-$\\
    \hline Friis Free Space & $-60.9$ & 2.00 & $\;\;-$ $\backslash$ $4.7$ \\
    \hline CI PG LoS - SC & $-60.9$ & 1.98 & $4.3$ $\backslash$ $4.8$\\
    \hline 3GPP TR 38.901 InF LoS & $-58.9$ & 2.15 & $4.3$ $\backslash$ $4.8$\\
    \hline \hline

    \multicolumn{4}{c}{Non Line-of-Sight} \\
    \hline \hline
    \multicolumn{1}{c|}{PG model} & \multicolumn{1}{c|}{$PG_{1m}$} & \multicolumn{1}{c|}{$n$} & \multicolumn{1}{c}{$\sigma$ $\backslash$ RMSE}\\
    \hline \hline
     
     
    \hline Measurement Fit (NLoS) &  $-39.6$ & 4.40 & $\!5.8$ $\backslash$ $\;\;-$\\
    \hline Friis Free Space & $-60.9$ & 2.00 & $\;\;\;-\;$ $\backslash$ $10.8$ \\
    \hline 3GPP TR 38.901 InF NLoS DL & $-47.1$ & 3.57 & $7.2$ $\backslash$ $6.5$\\
    \hline D. Chizhik \textit{et al.} & $-41.9$ & 4.04 & $\;\;-\;$ $\backslash$ $6.1$\\
    \hline \hline

    \end{tabular}

    }
    
    \vspace{3truemm}
    
\end{table}

Figs. \ref{PG}(a) and \ref{PG}(b) show the path gain in terms of the distance in the LoS and NLoS regions. Measurements correspond to both measurement campaigns after RSRP compensation according to eq. (2). Note that after RSRP compensation, the path gain is an inherent property of the scenario and does not depend on the chosen configuration. For this reason, data from both measurement campaigns are merged for the analysis.

For the LoS case, the measurement fit indicates a path loss exponent $n = 2.29$ and a standard deviation $\sigma = 4.6$ dB. This fit is compared with Friis free space model \cite{Friis}, close-in free space (CI) path-gain model for LoS in sparse clutter at 28~GHz  \cite{3GPP_CI}, and 3GPP TR 38.901 InF for LoS \cite{tr38901}. The path loss exponent close to two indicates a channel propagation dominated by the LoS path, resembling a free space channel on average. However, a noticeable high variability of $\sigma\nolinebreak=\nolinebreak4.6$~dB is found in the channel due to scattering in the sparse clutter, resulting in a multipath channel with constructive and destructive interference. Note that $\sigma$ indicates the goodness of fit on the data for which the model is designed, while RMSE indicates the goodness of fit between the model and the data in this work. An excellent agreement is found between the data and the models.

In the NLoS case, the measurement fit ($n = 4.40$, $\sigma\nolinebreak=\nolinebreak5.8$~dB) shows a scenario with high attenuation and distance dependant and high variability in the PG level. This is a direct consequence of the channel blockage, the received signal being the sum of the multipath channel contributions in the dense clutter machinery of the hall. This fit is compared with the Friis free space model, 3GPP TR 38.901 InF DL for NLoS \cite{tr38901}, and the model proposed in \cite{Valenzuela}. Friis model fails to fit the data due to the assumption of free space propagation. The latter two models agree well with the measurement fit, although they estimate a slightly lower path loss exponent. In addition, note that there is a sharp drop in the path gain beyond 30 m, due to the presence of a pallet rack whose height is higher than the BS height. Thus, this element resembles a wall which blocks the signal. This pallet rack is shown in the background of Fig. 1(c) and it is also illustrated in Fig. 4 as the southern horizontal region in the dense hall where the AMR is unable to navigate.

The fitness results are summarized in Table I, where it is shown that the RMSE values are around 4.7 dB for the LoS cases, and 6.6 dB for the NLoS cases. In both situations, the measured path gain exhibits a good agreement with the reference propagation models, which indicates that these models are a good choice for radio propagation prediction or coverage estimation in such industrial scenarios considering FR2 wall-mounted deployments with directional antennas. The main advantage of using a generic slope intercept model is that by means of only three parameters ($PG_{1m}$, $n$, $\sigma$) it is possible to accurately model the general behaviour of a heterogeneous industrial environment such as the one proposed in this work. Thus, the results can be easily extended to scenarios with similar characteristics. Conversely, for scenario-specific factors, it is appropriate to employ deterministic models \cite{modeling}.

\section{Beam Management Procedures}

The beam management concept is essential in 5G NR due to the use of several beams on both TX/RX sides. High gains from narrow beams ensure better communication performance in the TX-RX link. However, these narrow beams imply less spatial coverage, so they must be combined with management mechanisms that optimize the allocation of resources in the network in real-time, especially when considering mobile elements. In factory scenarios, the synchronization of AMRs in industrial processes is essential. That is why mechanisms such as beam sweeping, switching, and recovery are necessary. This Section analyzes some of the previous procedures given the two proposed TX configurations. A comprehensive understanding of these procedures in operational conditions is fundamental from the radio planning perspective. This is especially critical in factories, where the mobility of the agents involved, e.g. AMRs, is a factor to be taken into account. Therefore, the network must be able to quickly adjust the transmission beams to maintain a continuous and stable connection, aiming at a zero-interruption and optimizing the network resource allocation.

\subsection{Beam Recovery}

As previously stated, lower spatial coverage by the narrow beams implies a higher probability of beam failure in highly dynamic environments. The beam failure concept appears when the coverage provided by a beam cannot be maintained due to degradation of the TX/RX link for any reason, such as blocking or fading. To solve this issue, 5G NR performs a beam switching procedure to another beam which can guarantee good quality on the link \cite{beamforming, 5G_physical_layer}.

\begin{figure}[t]
    \centering
    \subfigure[]{\includegraphics[width= 0.99\columnwidth]{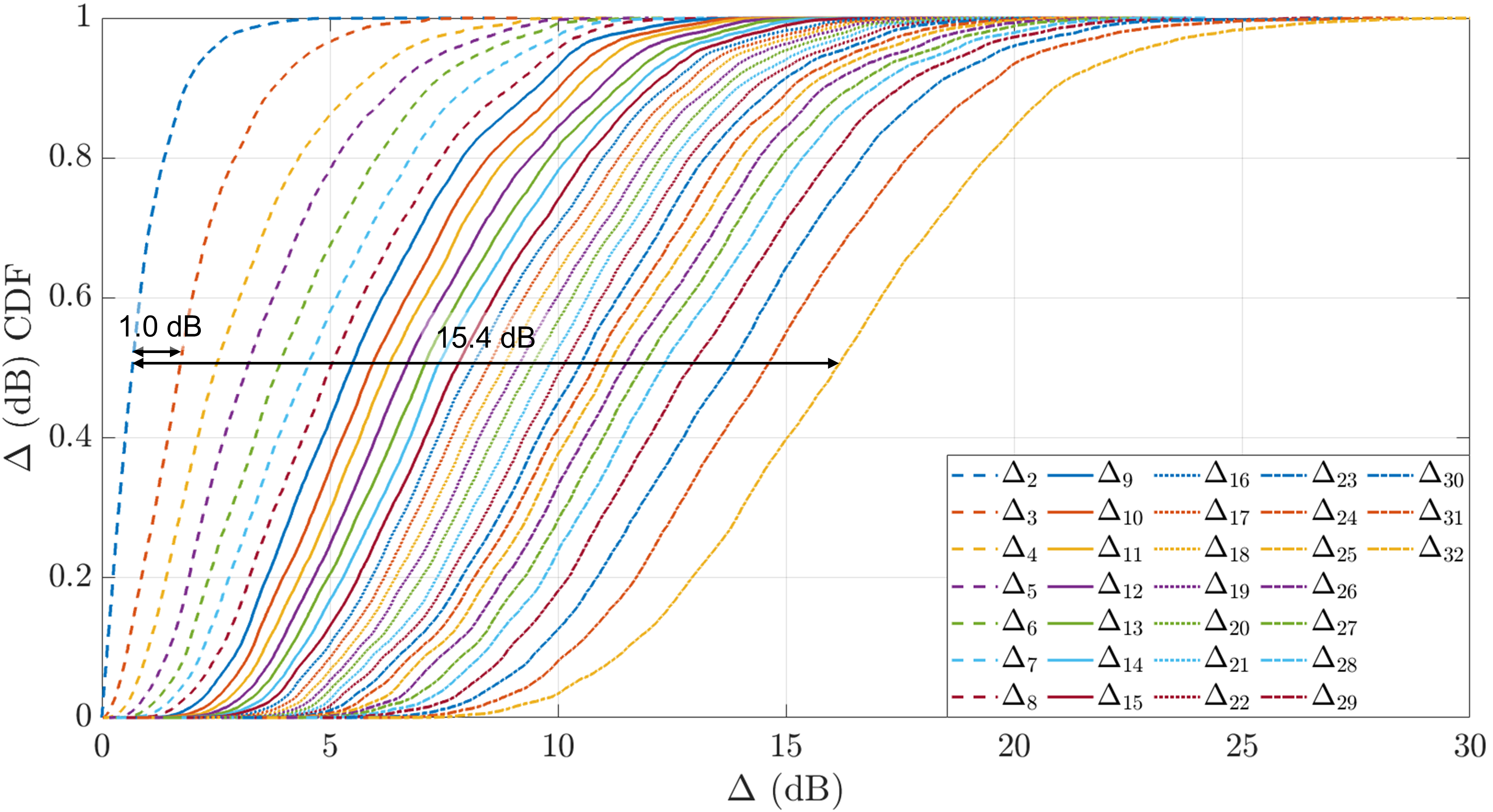}
	}
	\subfigure[]{\includegraphics[width= 0.99\columnwidth]{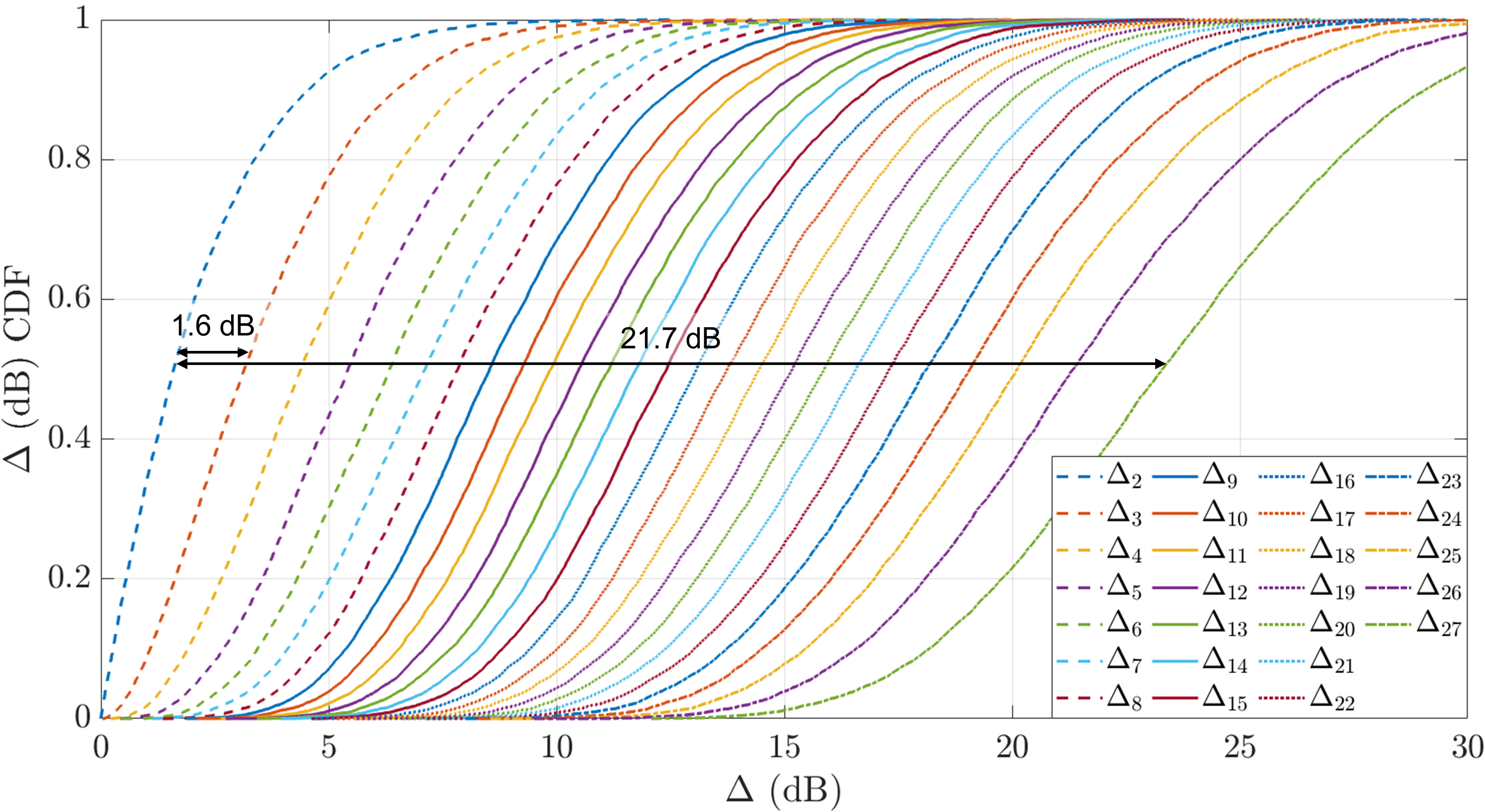}
	}
    \caption{$\Delta_{i}$ CDF for (a) configuration A and (b) configuration B.}
	\label{BFR}
\end{figure}

In order to ensure the success of the beam failure recovery procedure, the new beam to which the user connects should provide the same or similar quality as prior to beam failure. Assuming that the user is served by the beam with the strongest RSRP, the quality of the beam recovery procedure can be quantified as $\Delta_{i} = \textrm{RSRP}_{1^{st}} - \textrm{RSRP}_{i^{th}}$, where $\textrm{RSRP}_{i^{th}}$ is the $i$-th strongest beam given a specific location in the scenario. Therefore, $\Delta_{i}$ specifies the RSRP difference between the strongest and the $i$-th strongest beam in case a beam failure recovery procedure is required. Values tending to zero would indicate that the channel would barely be degraded due to this procedure. Figs. \ref{BFR}(a) and \ref{BFR}(b) show the CDF for $\Delta_{i}$ values in both measurement campaigns. Two main differences are found: (i) Configuration A shows CDF curves closer to zero compared to the second module configuration. Specifically, $\Delta_{2}$ median value is only 0.7~dB, which means a degradation of the TX-RX link below 1~dB if a switching procedure is performed from the strongest to the second strongest beam. However, $\Delta_{2}$ median for configuration B is 1.6 dB. (ii) Distance between adjacent $\Delta_{i}$ curves is smaller in configuration A. For example, $\Delta_3 - \Delta_2$ is 1.0 dB for configuration A, while this value goes to 1.6 dB for configuration B when $P(\Delta_i < \textrm{50\%})$. This leads to a mitigation of power loss during the beam switching process in configuration~A when the serving beam is unavailable due to, for instance, a Line-of-Sight blockage. Specifically, the average median distance between two consecutive curves is 0.5 dB and 0.9 dB, for configurations A and B, respectively.

\renewcommand{\arraystretch}{1.5}
\setlength{\tabcolsep}{4pt}

    \begin{table}
    \centering
    \caption {$\Delta$ values for several probabilities in both configurations} \label{tabla_3} 

    \begin{tabular}{c|c|c|c||c|c|c}
    \cline{2-7}
    \multicolumn{1}{c|}{} & \multicolumn{3}{c||}{ Configuration A} & \multicolumn{3}{c}{Configuration B}\\
    \hline \hline
     
     
     $P(\Delta_i < x)$ & $\Delta_{2}$ & $\Delta_{3}$ & $\Delta_{4}$ & $\Delta_{2}$ & $\Delta_{3}$ & $\Delta_{4}$\\
    \hline \hline 25\% & 0.3 dB & 1.0 dB & 1.6 dB & 0.7 dB & 1.9 dB & 3.0 dB\\
    \hline 50\% & 0.7 dB & 1.7 dB & 2.5 dB & 1.6 dB & 3.2 dB & 4.4 dB\\
    \hline 75\% & 1.2 dB & 2.6 dB & 3.9 dB & 3.0 dB & 4.8 dB & 6.2 dB\\
    \hline \hline

    \end{tabular}

    
\end{table}

The two previous differences are mainly due to the configuration of beams previously described in Fig. 3. In configuration A, a higher number of available beams as well as a higher density of beams in the spatial domain lead to better $\Delta$ values in the case of requiring a beam recovery procedure. Configuration B distributes the beams in the spatial domain, achieving a better global coverage for the strongest beam, as shown in Fig. 4, but this higher spatial separation implies a higher $\Delta$ when beam switching is required. Table III summarizes the three strongest backup beams, i.e., $\Delta_{2}$, $\Delta_{3}$ and $\Delta_{4}$ for both configurations and multiple percentiles $P(\Delta_i < x)$. In communications where the decrease of RSRP in case of beam failure is not critical, configuration B is suitable as long as the strongest beam is available, as shown in the overall coverage in Fig. \ref{RSRP_map}. However, if it is necessary to maintain a constant RSRP, configuration A is more appropriate as the beams are grouped in a smaller spatial region. This fact is reflected in $\Delta_i$ values, which are minimized for configuration A, as illustrated in Fig. \ref{BFR} and Table III. In summary, the results presented in this subsection aim to characterize the channel from a physical layer perspective, considering a specific spatial distribution of the beams located at the transmitter side. A first analysis of these results given a specific configuration may lead to an appropriate choice of beam tracking or beam selection algorithms at higher protocol layers.

\subsection{Beam sweeping and switching}

\begin{figure}[t]
    \centering
    \subfigure[]{\includegraphics[width= 0.39\columnwidth]{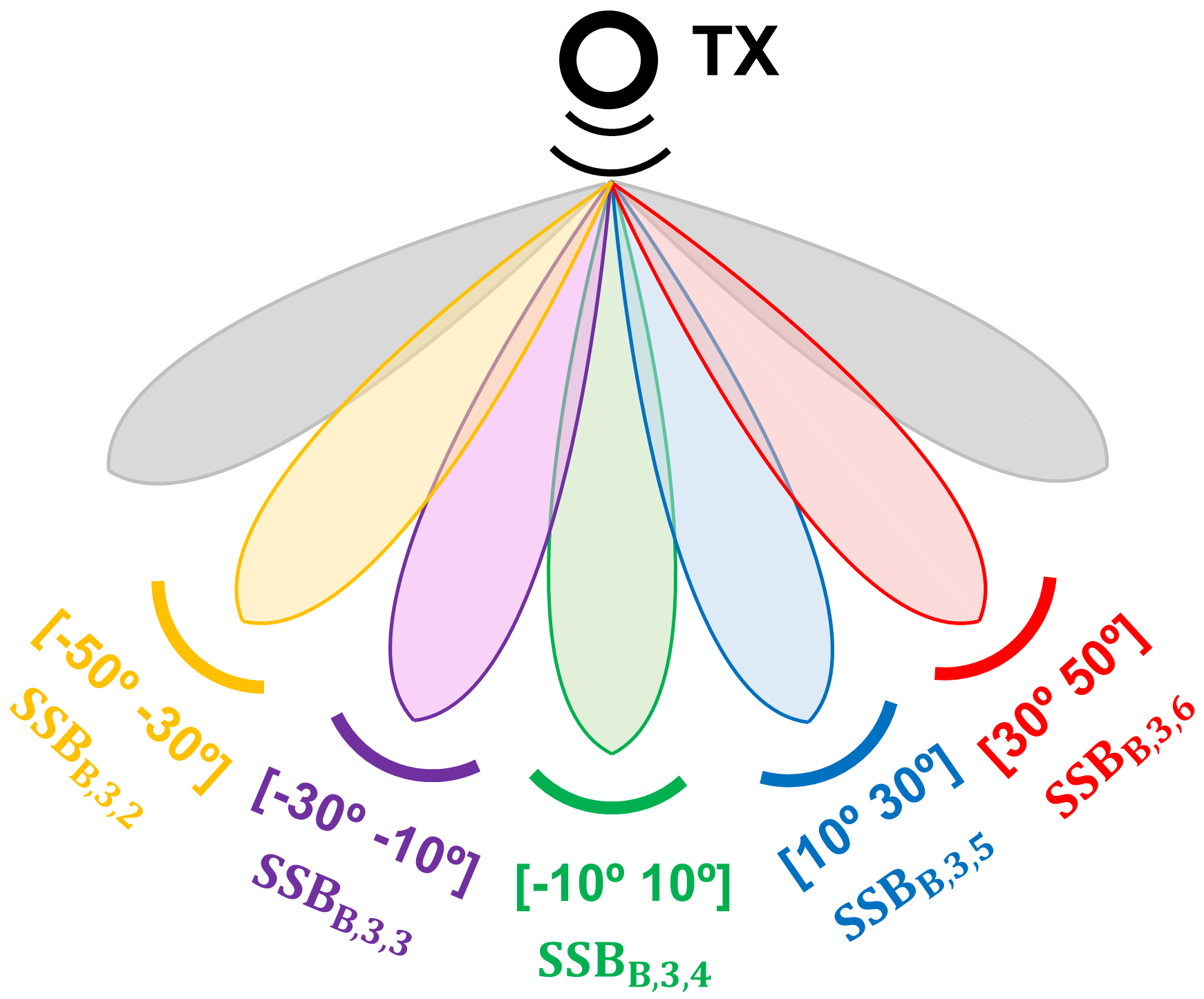}
	}
	\subfigure[]{\includegraphics[width= 0.57\columnwidth]{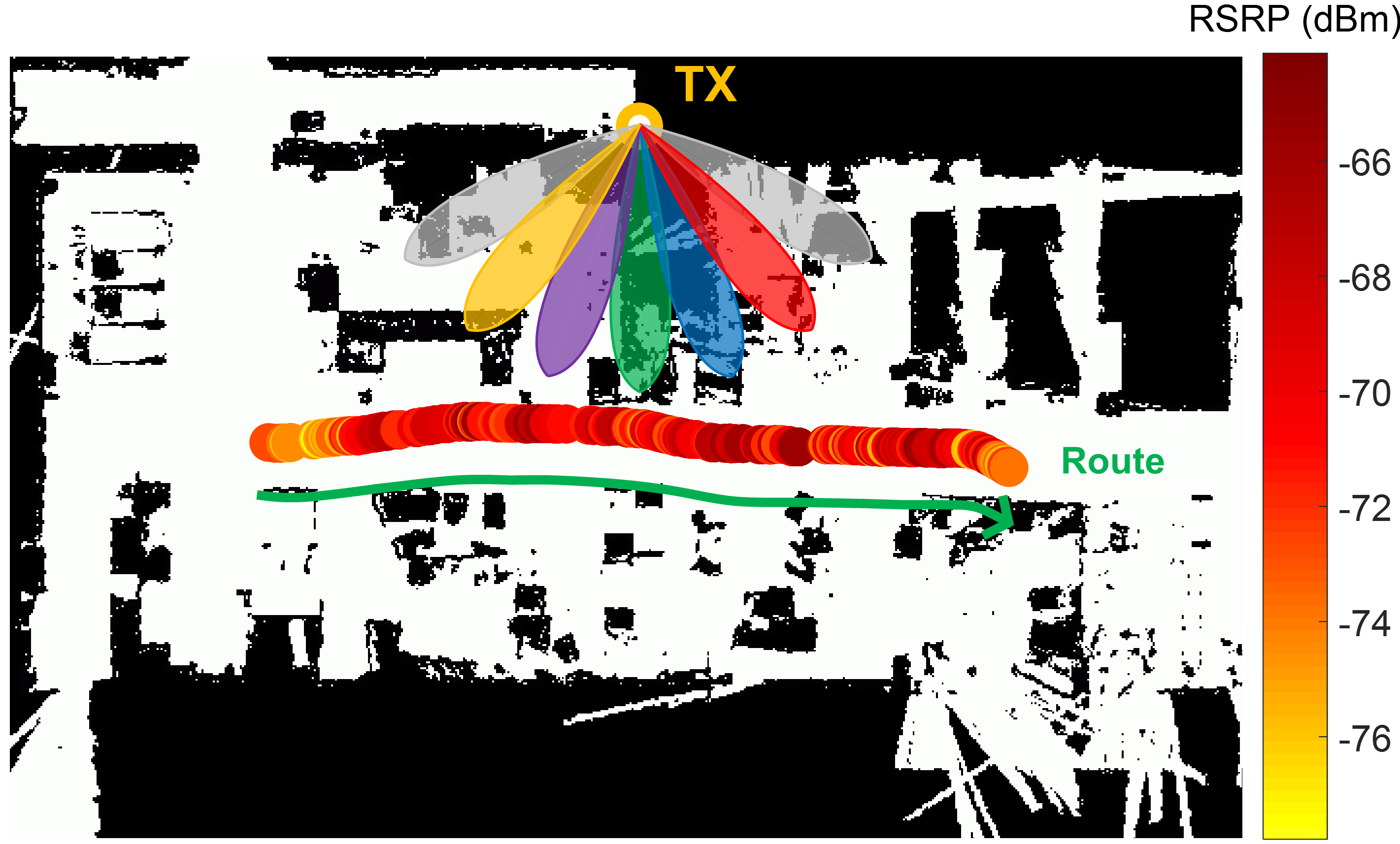}
	}
	\subfigure[]{\includegraphics[width= 1\columnwidth]{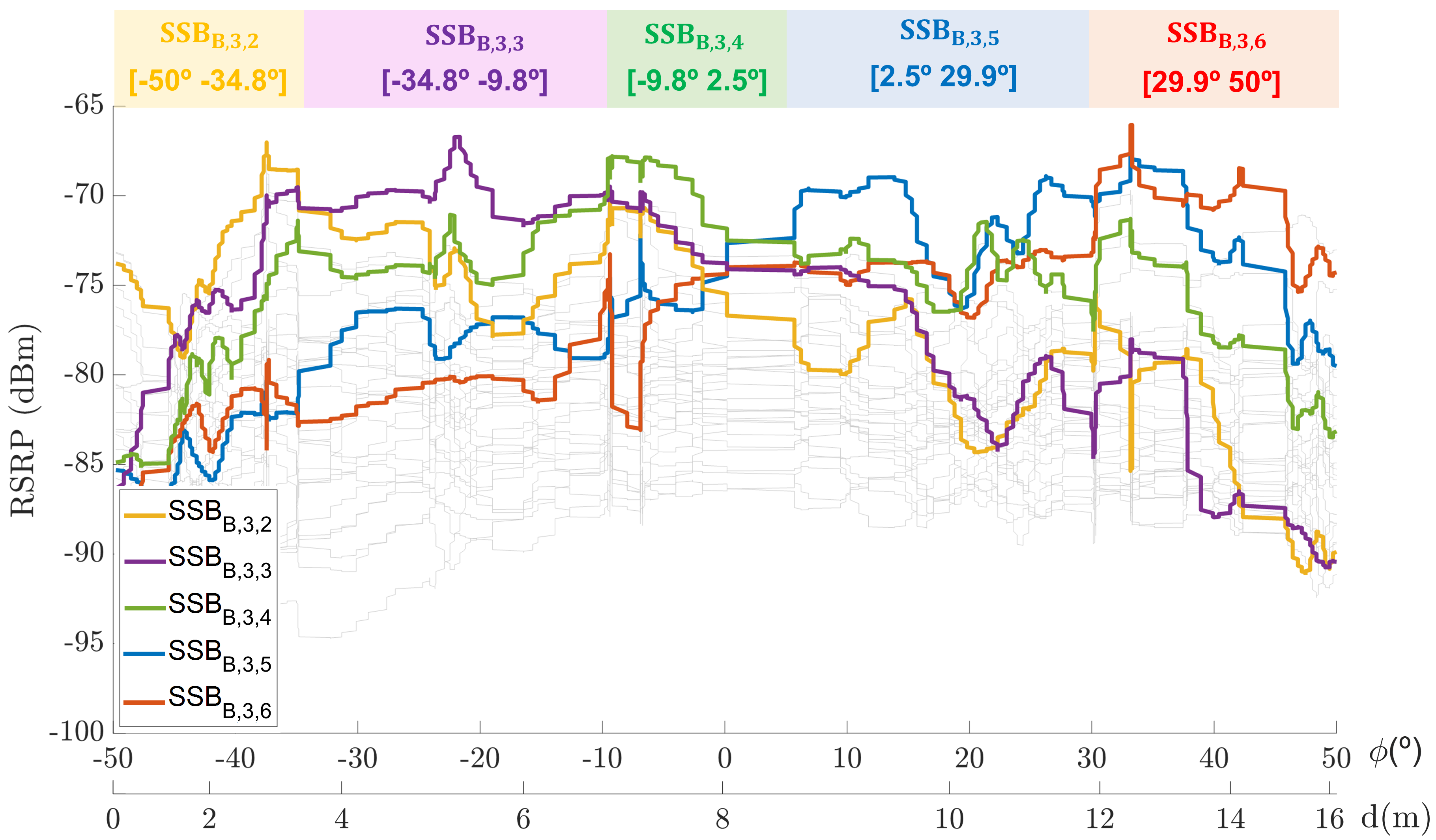}
	}
    \subfigure[]{\includegraphics[width= 1\columnwidth]{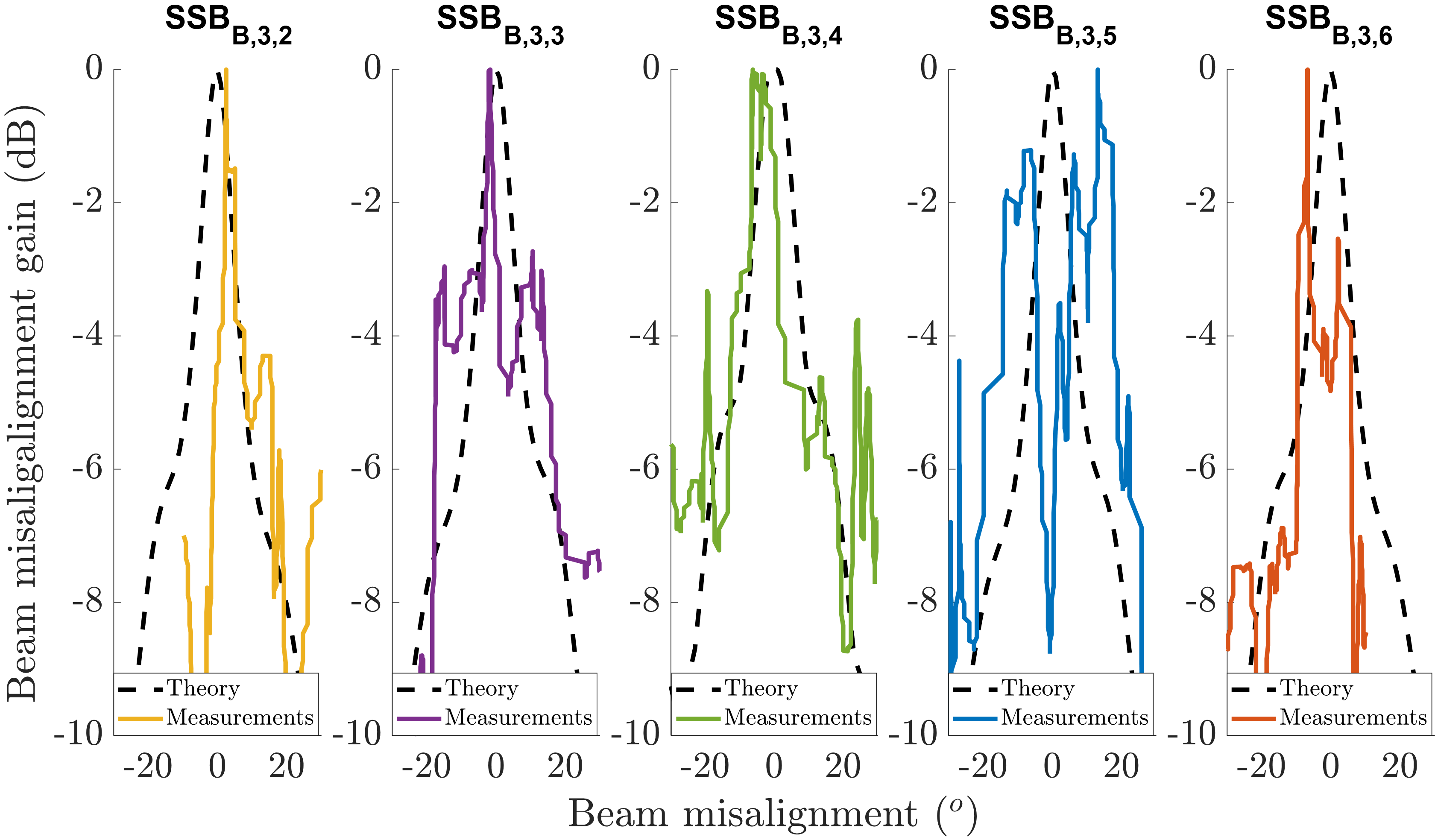}
	}
    \caption{(a) Azimuth beam sweeping diagram for the bottom row of beams in the configuration B, (b) route followed by the AMR in the LoS hall, (c) angular distribution of the RSRP for every SSB and (d) theoretical and experimental measurements of the antenna gain for 5 SSBs.} 
	\label{angular_analysis}
\end{figure}

In addition to the beam recovery procedure, beam switching scheduling is essential for the optimal management of signaling resources on the TX side \cite{beamforming, 5G_physical_layer}. If a serving beam pattern is found based on the location in the industrial environment, a priori scheduling for the beam assignment can be made. While in free space, this scheduling would be obvious based on the antenna radiation pattern, in an industrial environment such as the one studied, this scheduling could differ due to scattering in the environment. In order to analyze the effect of the propagation channel on the radiation pattern, a 16-meter route is configured with the AMR along the main corridor of the hall with LoS. This corridor is chosen since it is considered the operating area according to the RSRP level and cell radii analyzed in previous sections. This route, which covers azimuth angles from $-50\degree$ to $50\degree$, coincides with the boresight of the following beams in configuration~B:  $\textrm{SSB}_{B,3,2}$, $\textrm{SSB}_{B,3,3}$, $\textrm{SSB}_{B,3,4}$, $\textrm{SSB}_{B,3,5}$ and $\textrm{SSB}_{B,3,6}$. Fig.~\ref{angular_analysis}(a) depicts these five SSBs in the lowest row of the beam pattern, as illustrated in Fig. \ref{patterns}(b). The angular ranges denote the spatial region in which the beam gain is higher than the others. Fig. \ref{angular_analysis}(b) shows the specific route followed by the AMR and the maximum RSRP received at each location. Fig. \ref{angular_analysis}(c) presents the RSRP for every SSB at each azimuth angle in the route after removing the small-scale fading effect. This has been performed by averaging the acquired measurements in the route over 40 wavelengths \cite{Lee_criteria, congreso_troels}. A second axis is included with the distance traveled by the AMR. Note that the relationship is not linear with $\phi$ since the distance between the AMR and the TX is not constant along the route. The RSRPs from beams shown in Fig. \ref{angular_analysis}(a) are highlighted. The angular ranges in which these beams predominate are denoted at the top. It is shown a good agreement between the beams predicted in Fig. \ref{angular_analysis}(a) by the radiation pattern, and the strongest beams along the angular domain of the route. Finally, Fig. \ref{angular_analysis}(d) shows a comparison between the theoretical and the experimental gain of each analyzed SSB. Due to the strong multipath effect of the industrial environment, although each SSB predominates in the expected angular region as illustrated in Fig. \ref{angular_analysis}(c), there
is a certain variability of the gain along the angles due to the in-phase or out-of-phase summation of the multipath components. 

\begin{figure*}[!b]
    \centering
    \subfigure[]{\includegraphics[width= 0.313\textwidth]{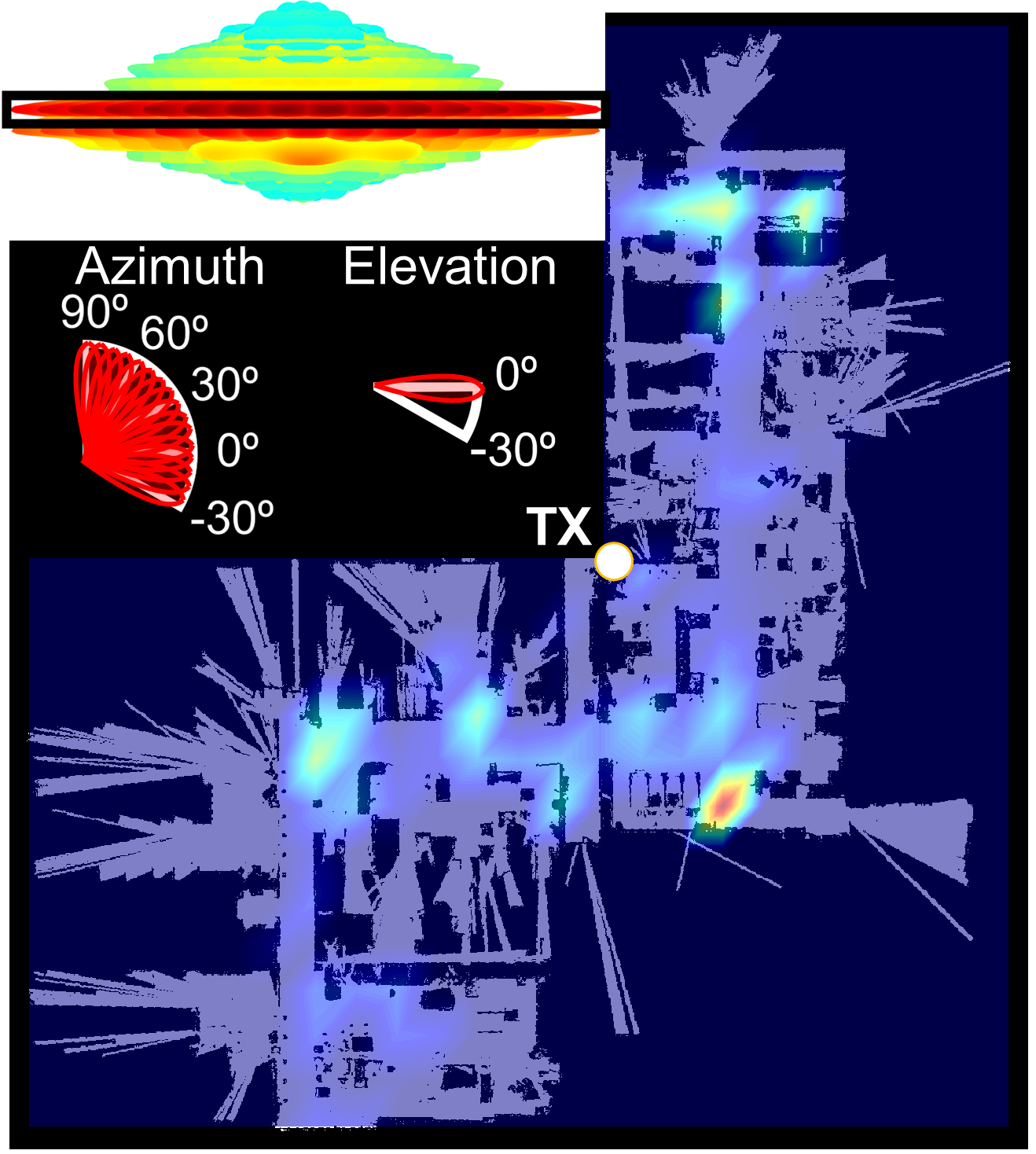}
	}
	\subfigure[]{\includegraphics[width= 0.313\textwidth]{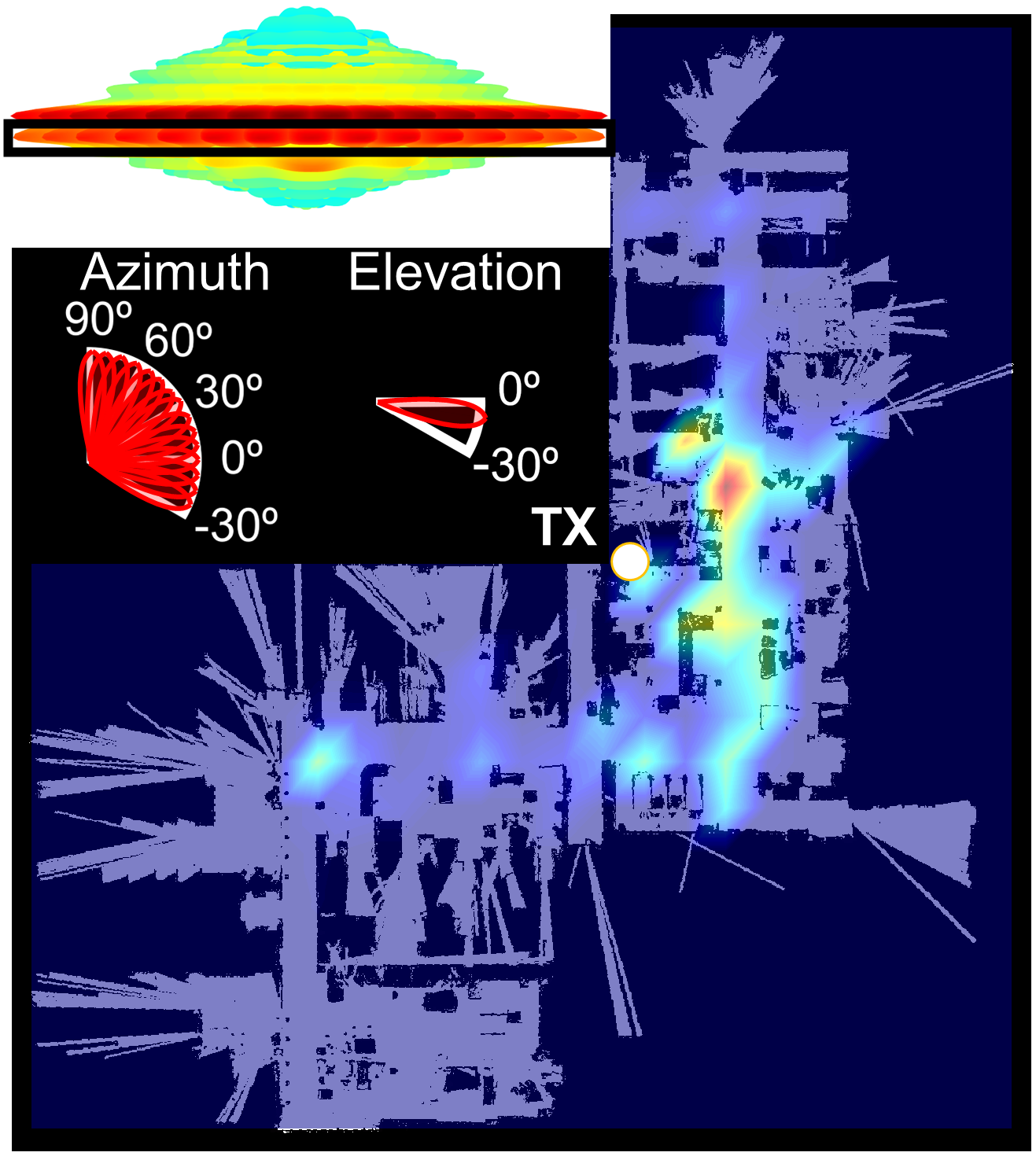}
	}
	\subfigure[]{\includegraphics[width= 0.313\textwidth]{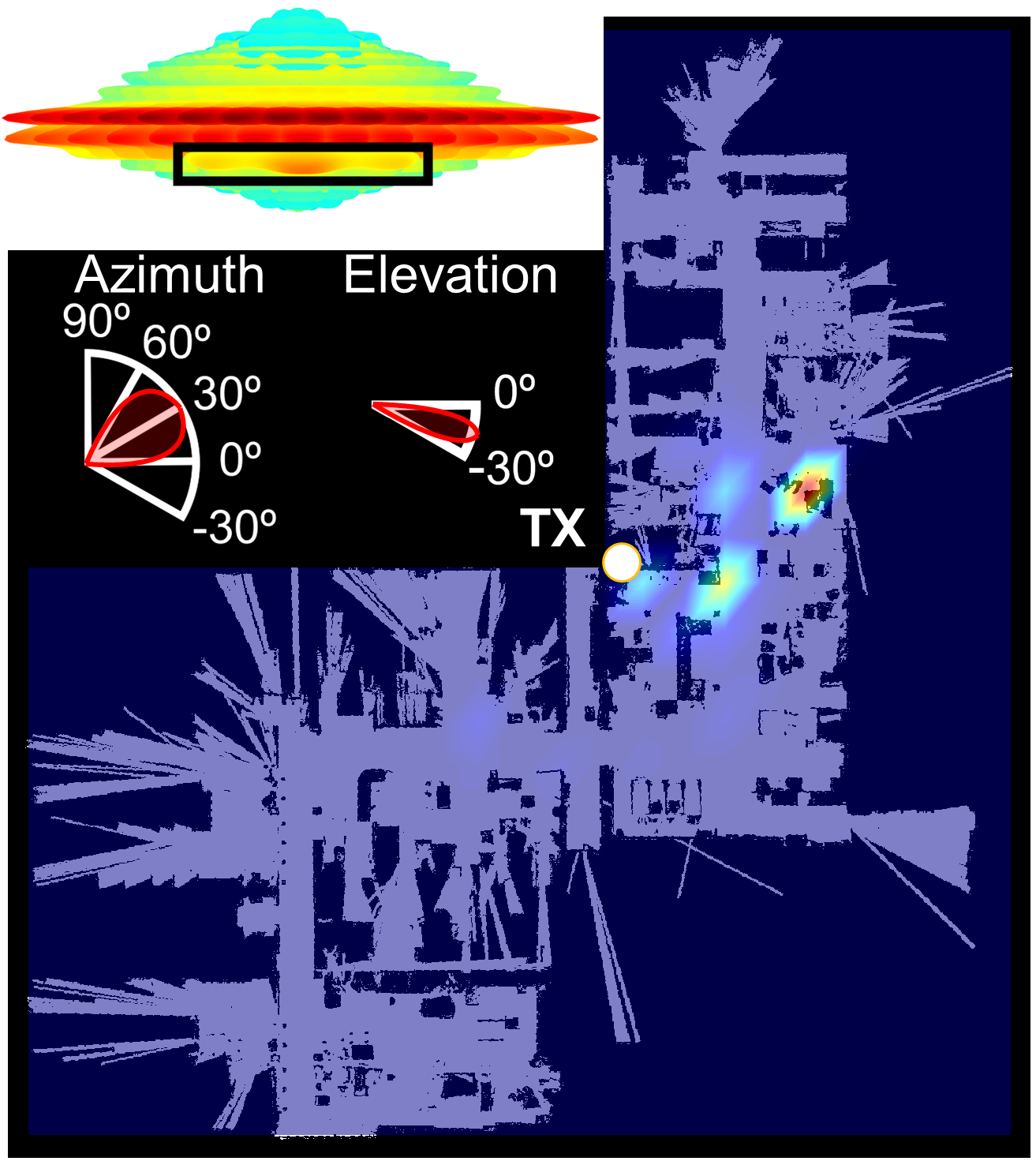}
	}
	
	 \subfigure[]{\includegraphics[width= 0.23\textwidth]{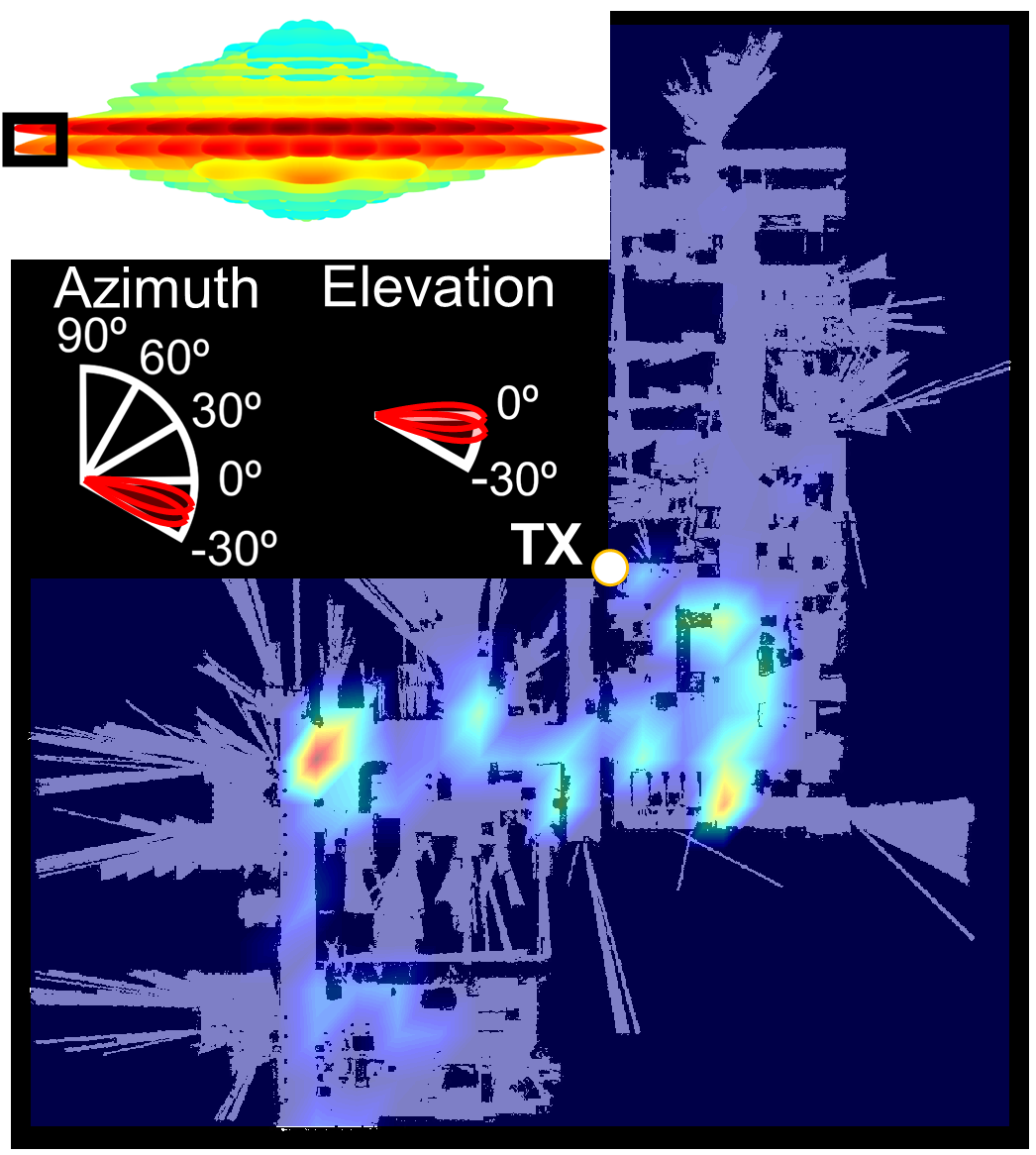}
	}
	\subfigure[]{\includegraphics[width= 0.23\textwidth]{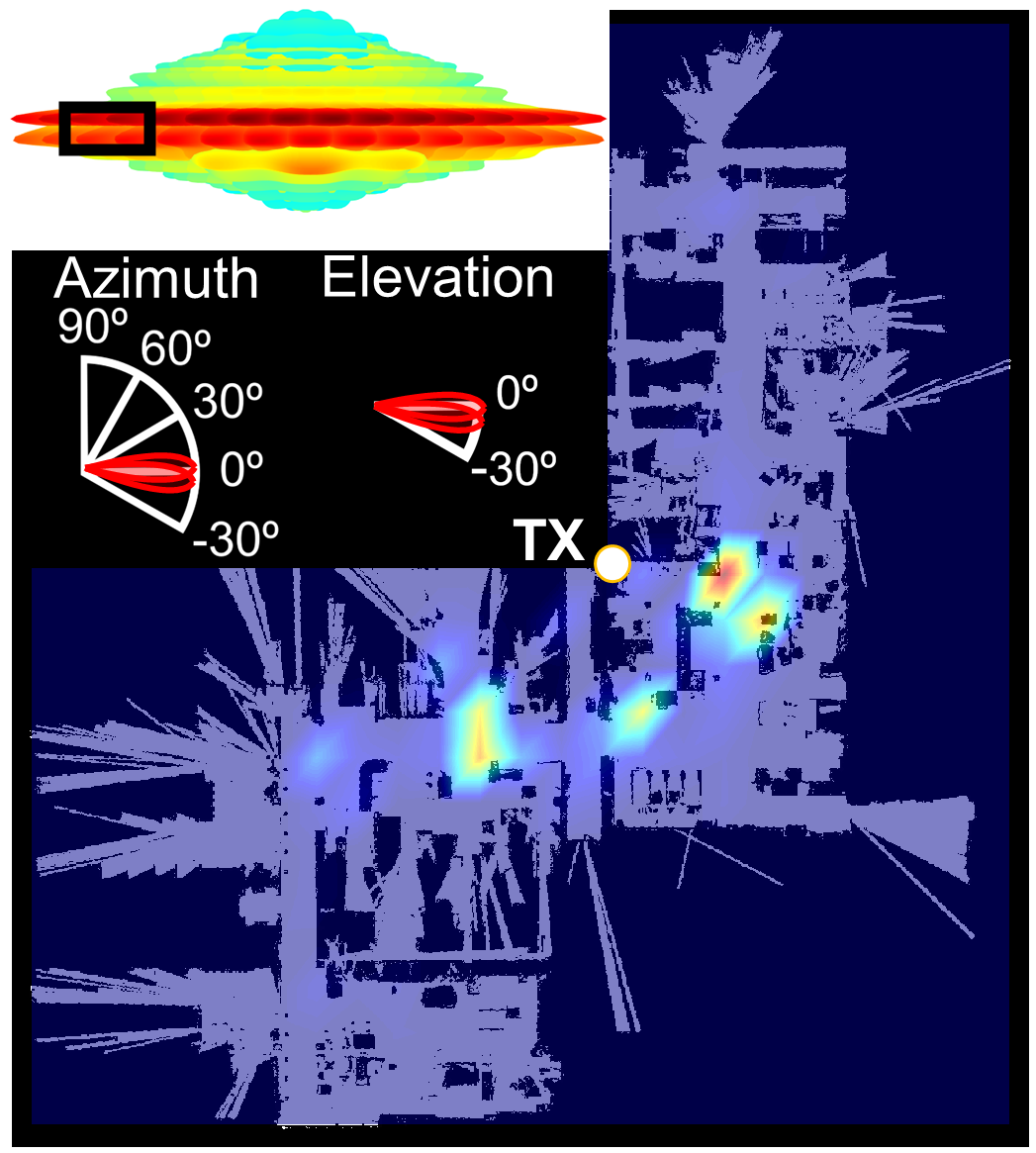}
	}
	\subfigure[]{\includegraphics[width= 0.23\textwidth]{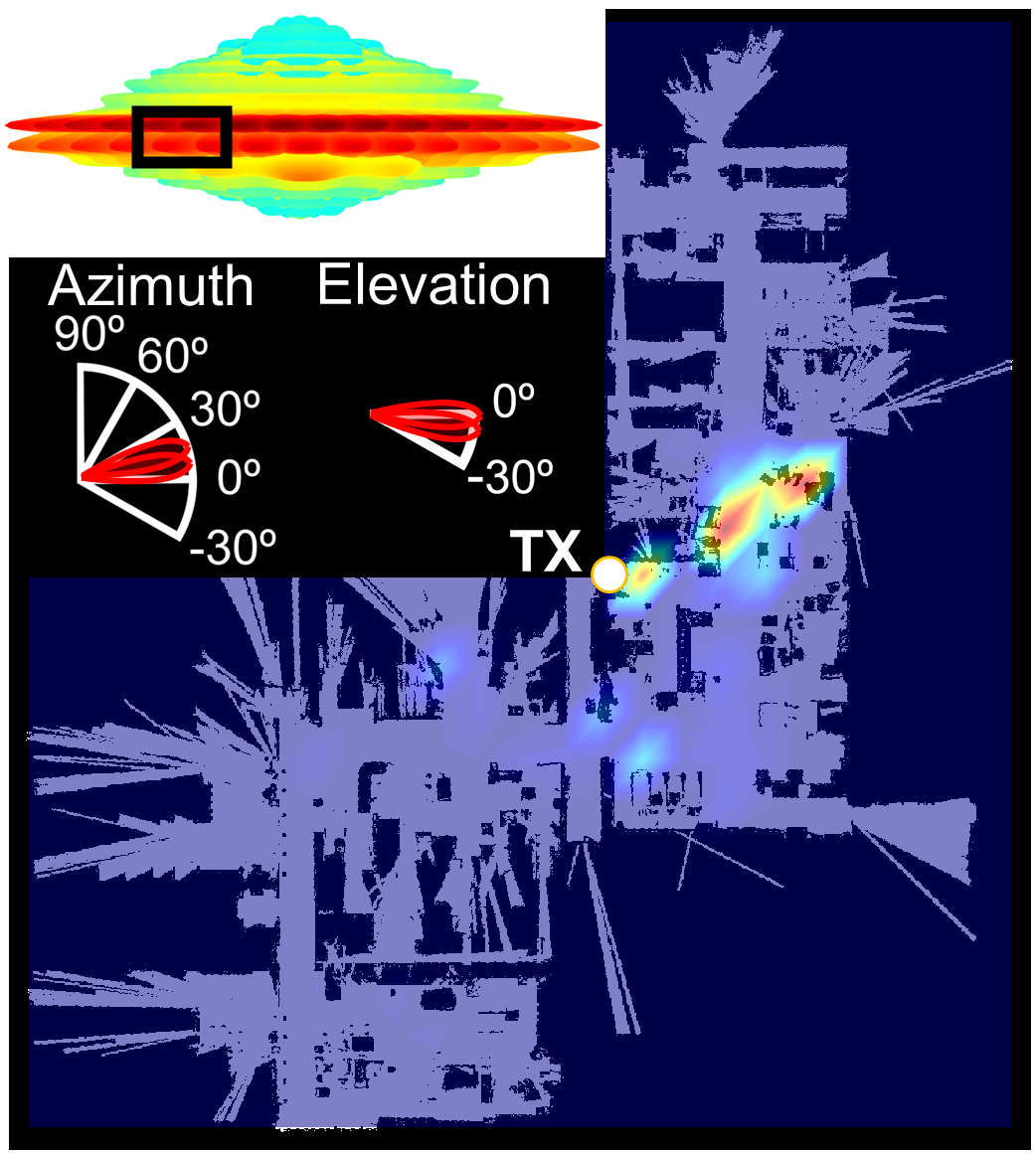}
	}
	\subfigure[]{\includegraphics[width= 0.23\textwidth]{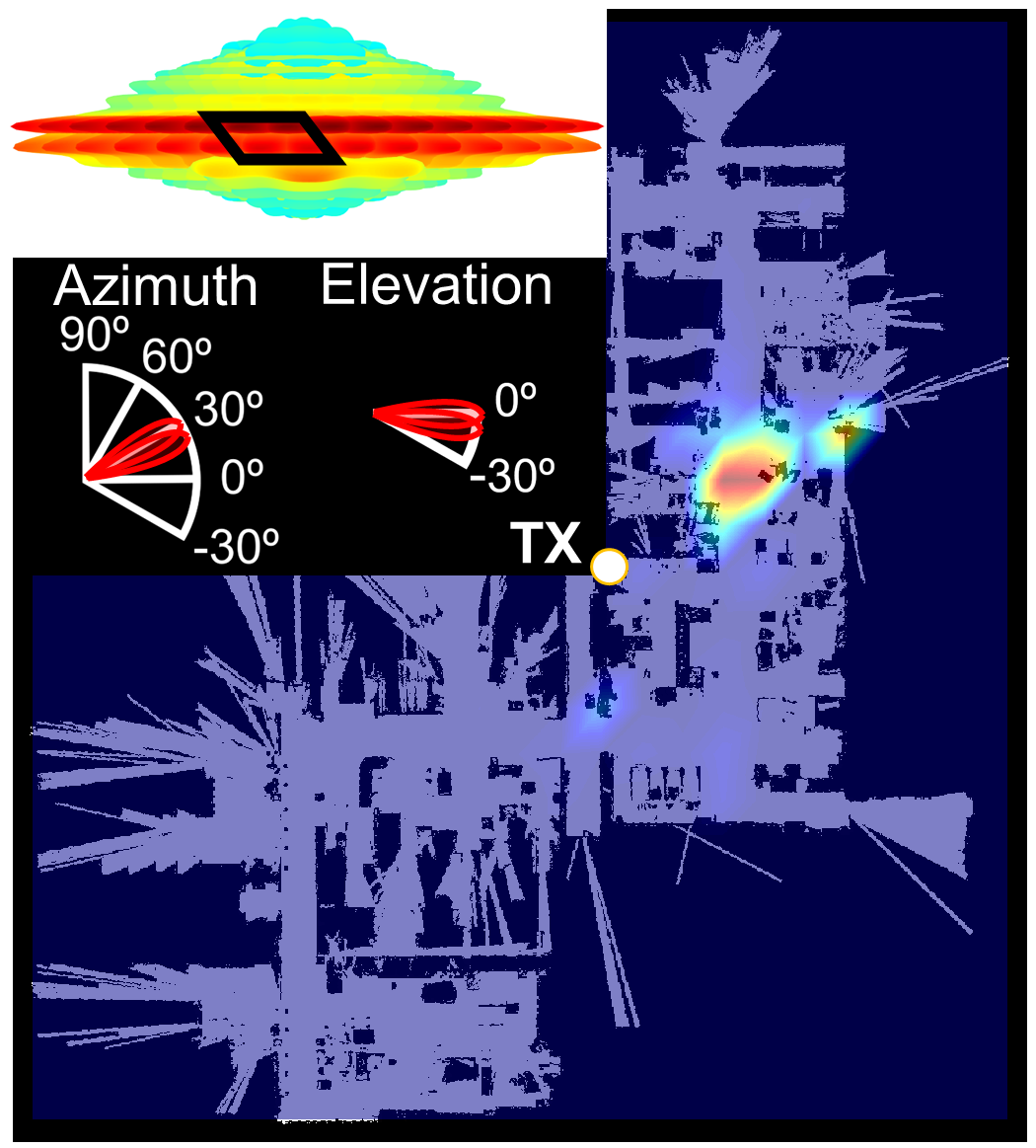}
	}
	
	 \subfigure[]{\includegraphics[width= 0.23\textwidth]{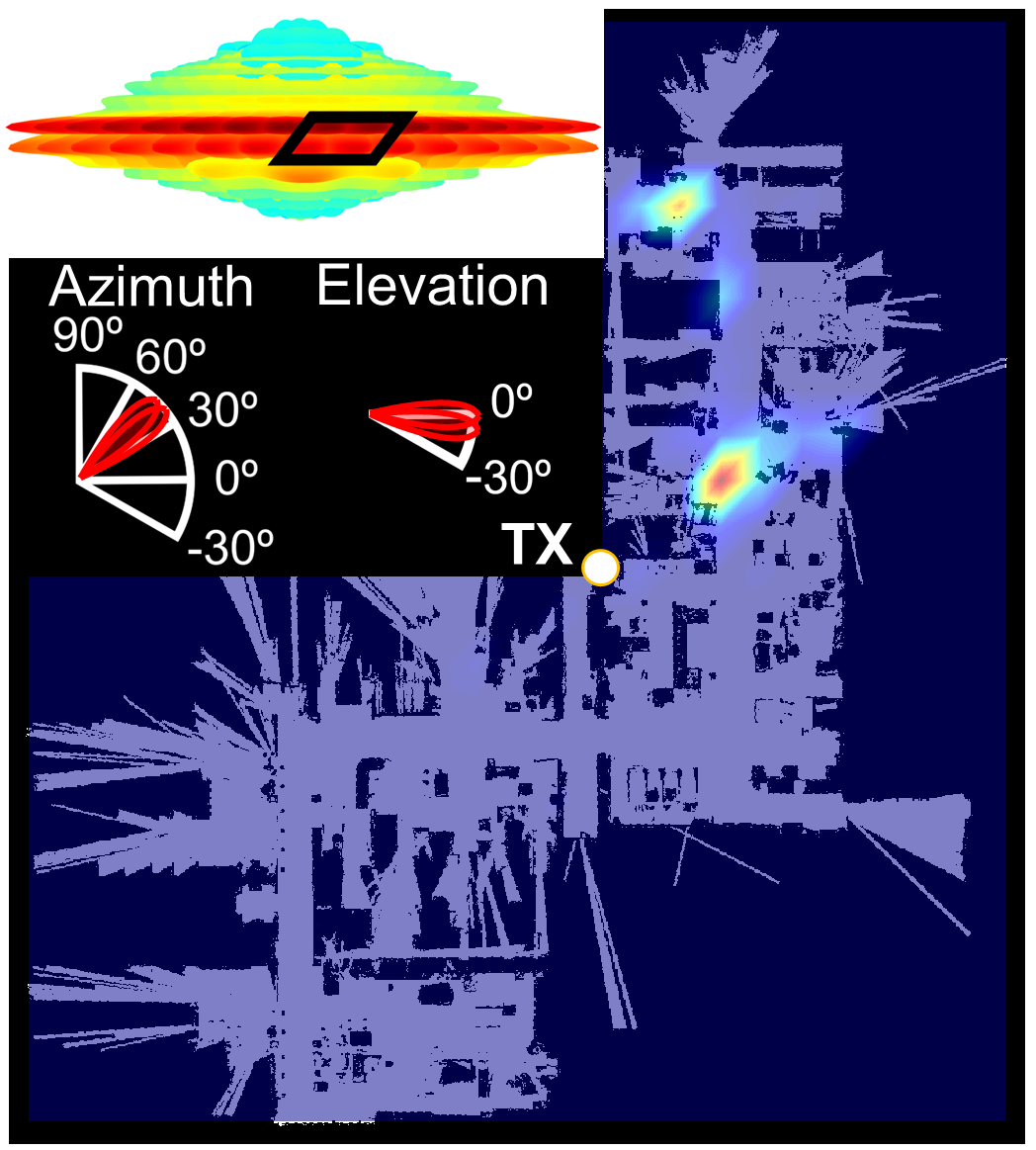}
	}
	\subfigure[]{\includegraphics[width= 0.23\textwidth]{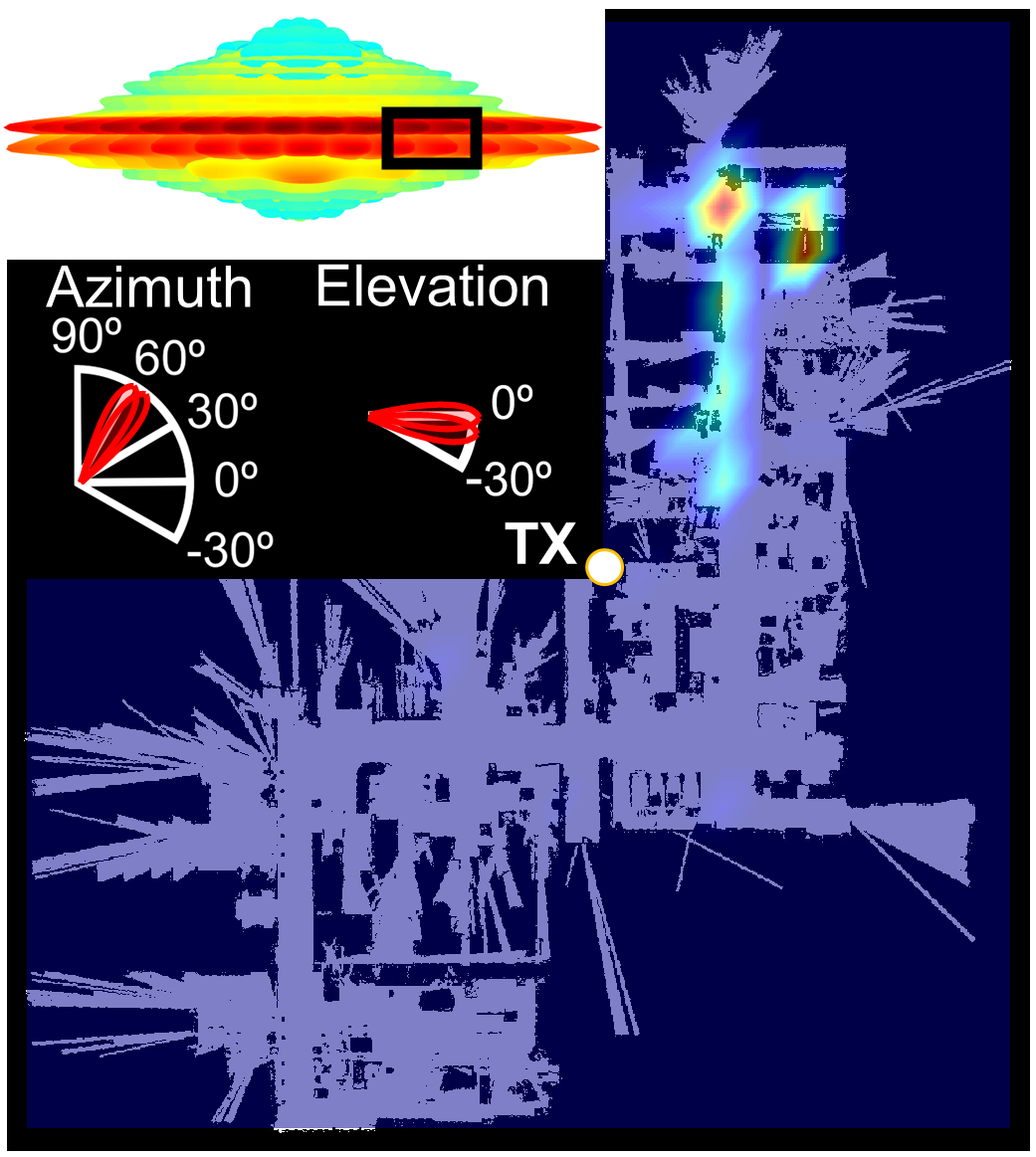}
	}
	\subfigure[]{\includegraphics[width= 0.23\textwidth]{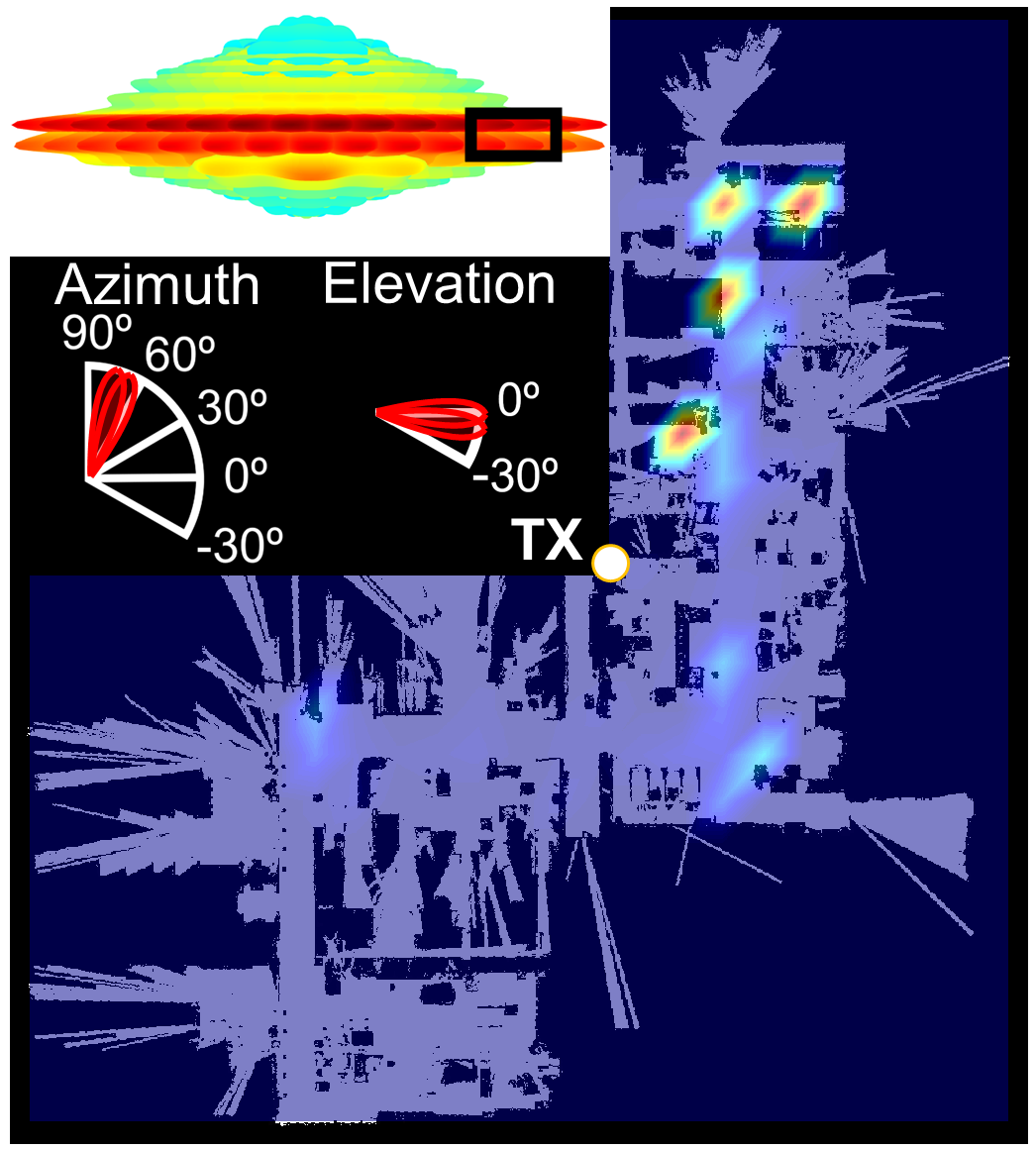}
	}
	\subfigure[]{\includegraphics[width= 0.23\textwidth]{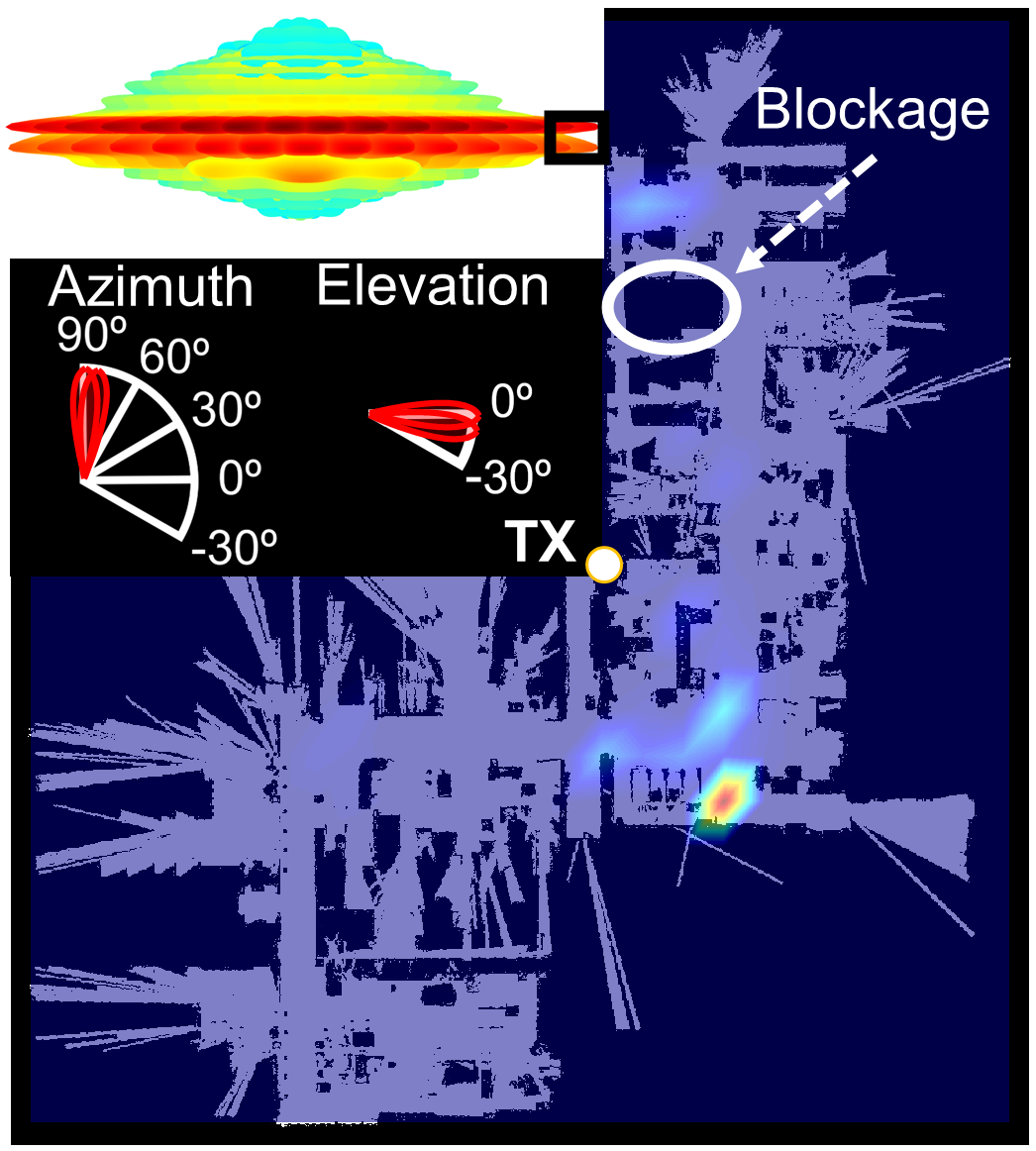}
	}
	
    \caption{Coverage maps of the factory in terms of SSB clusters: (a)-(c) separate the SSBs in three elevation angles and (d)-(k) consider eight azimuth regions.} 
	\label{coverage_maps}
\end{figure*}

The results shown in this Section demonstrate that in the LoS zone, the direct ray generally prevails even with the strong scattering and the large-scale fading expected in an industrial environment. Therefore, in order to decrease the signaling in the network, scheduling of the beam assignment based on the AMR position can be considered if the route is known. Nevertheless, it must be noted that this multipath environment introduces some variance of the signal level in the angular domain. Thus, for multipath scenarios whose high variability in signal level overshadows the LoS due to constructive and destructive interference, the need for algorithms to mitigate signal degradation is remarkable.

\begin{figure*}[!b]
    \centering
    \subfigure[]{\includegraphics[width= 0.320\textwidth]{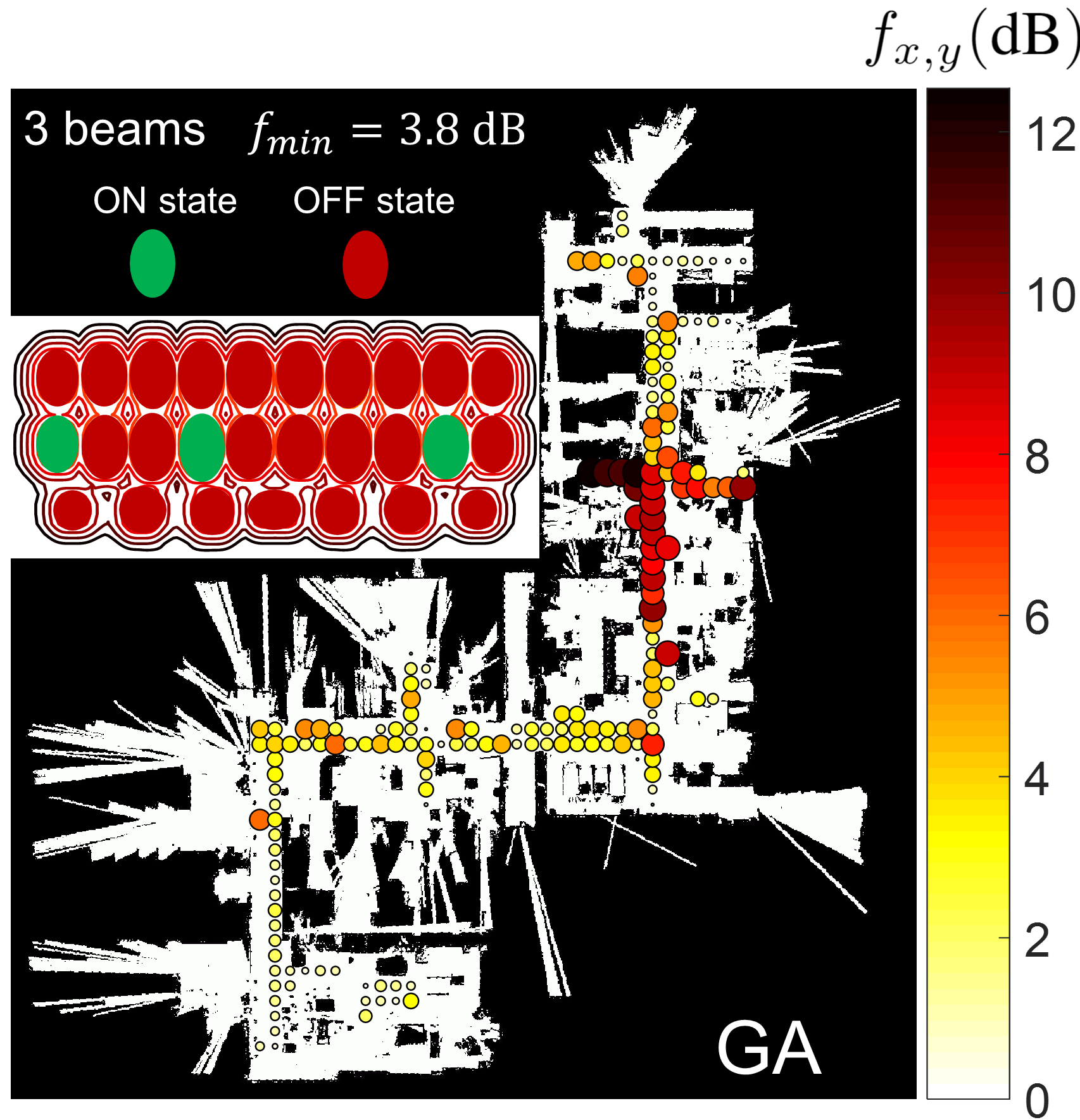}
	}
	\subfigure[]{\includegraphics[width= 0.320\textwidth]{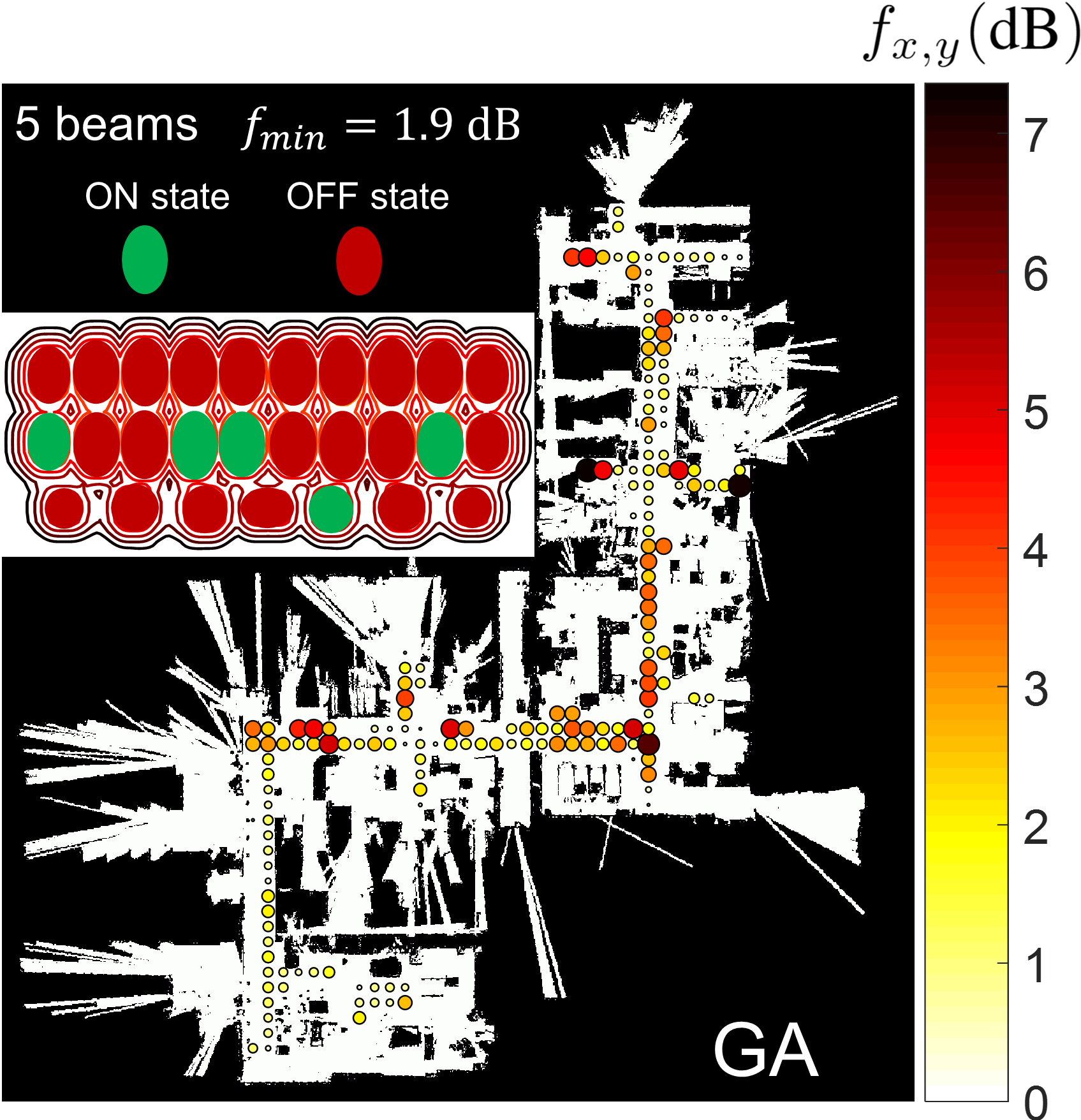}
	}
	\subfigure[]{\includegraphics[width= 0.320\textwidth]{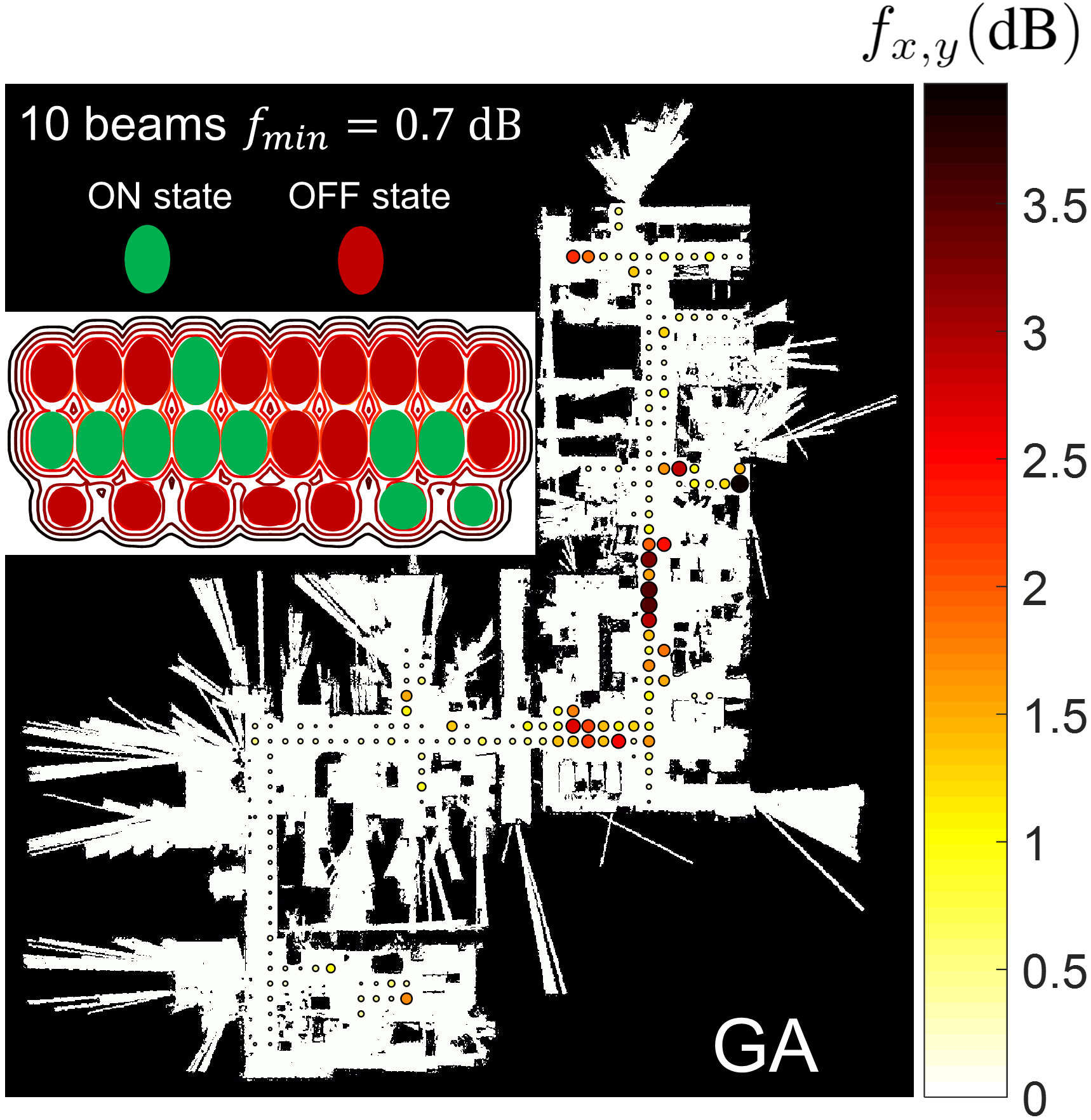}
	}
 	\subfigure[]{\includegraphics[width= 0.320\textwidth]{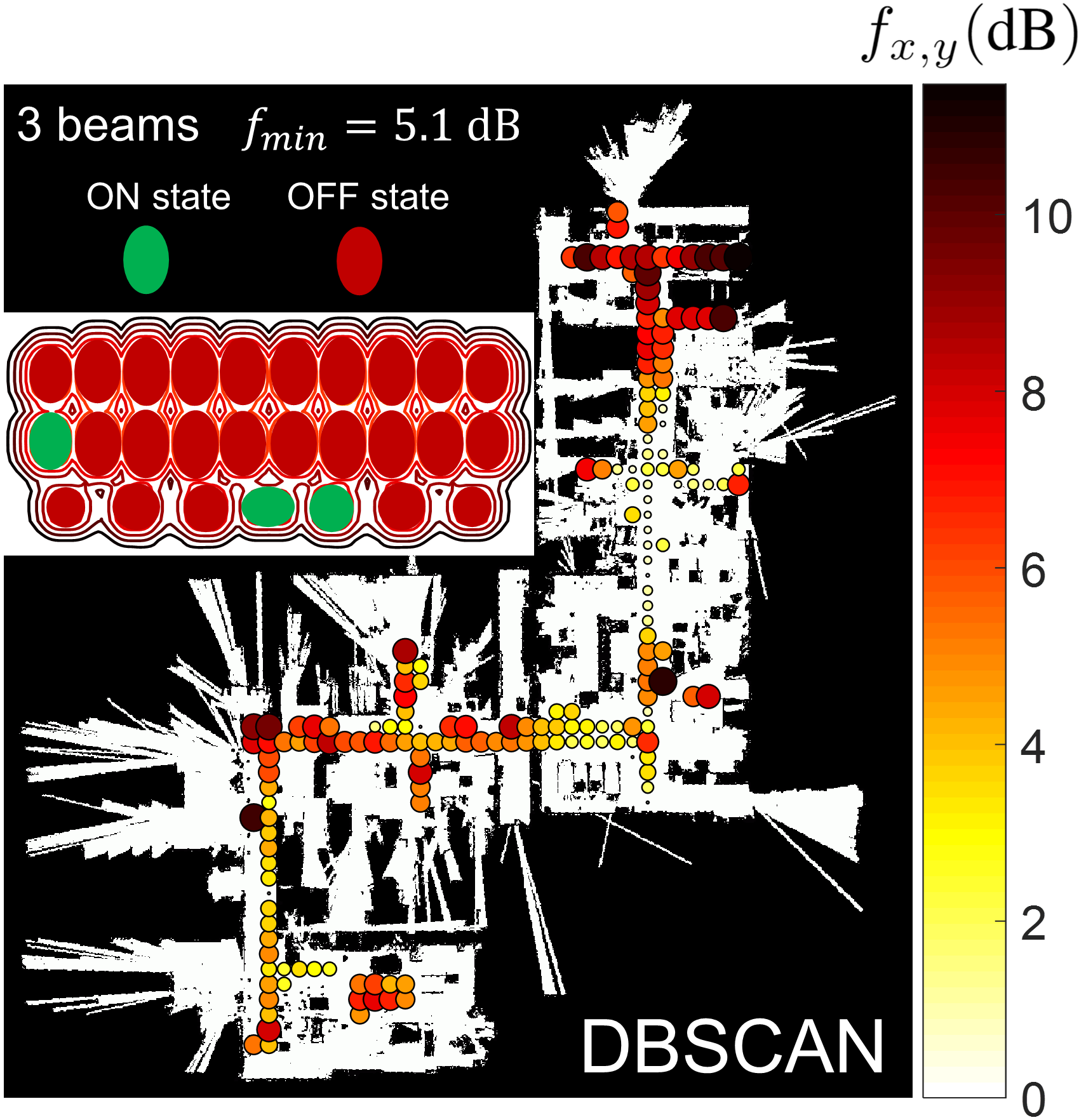}
	}
 	\subfigure[]{\includegraphics[width= 0.320\textwidth]{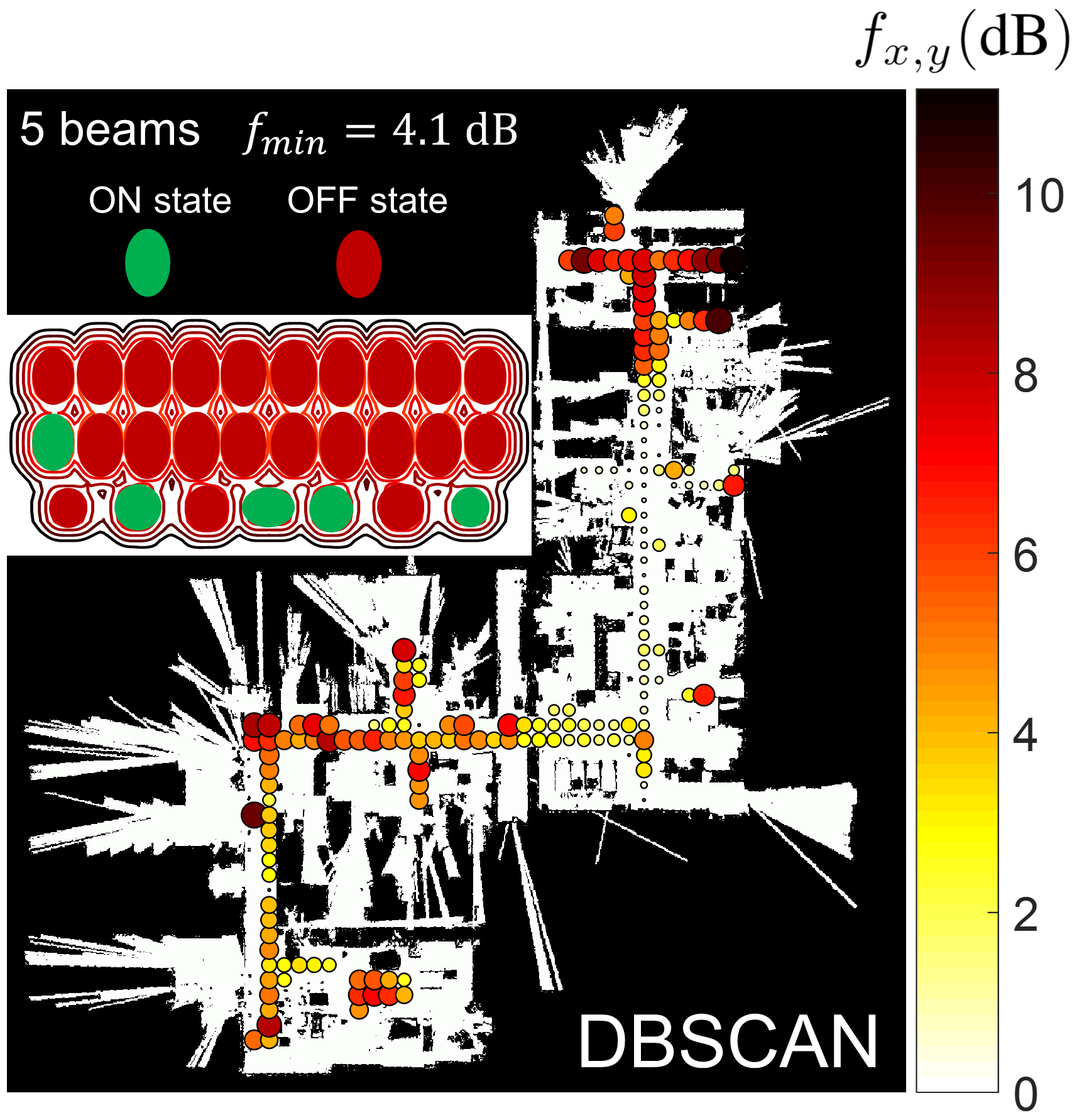}
	}
 	\subfigure[]{\includegraphics[width= 0.320\textwidth]{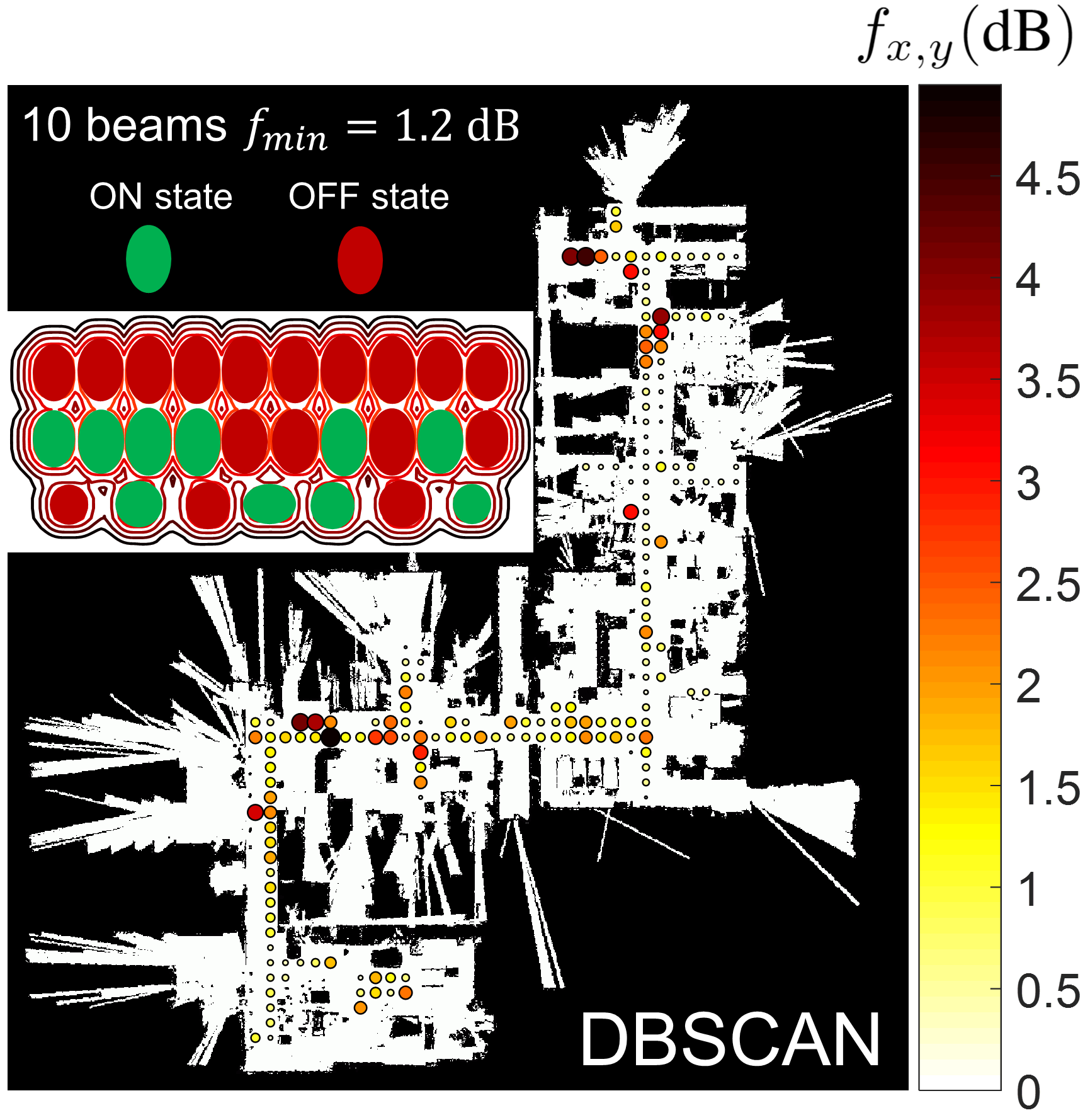}
	}
    \caption{Solution of the genetic algorithm (a)-(c) and DBSCAN clustering (d)-(f) for the beam switch-off problem with $\xi = 3$, $\xi = 5$ and $\xi = 10$.} 
	\label{optimization}
\end{figure*}

\section{Beam-specific coverage and Switch-off Optimization study}

As mentioned in previous Sections, 5G NR support for beamforming is intended to improve the spatial coverage by taking advantage of the highly directive beams provided in the mmWave/FR2 band. To check the effectiveness of optimal spatial coverage at the factory, this section analyzes some coverage maps provided by the beams involved in the beam sweeping procedure for the first measurement campaign. These maps are calculated by dividing the factory into a grid with inner areas whose dimensions are $2.7 \times 2.4 \textrm{ m}^2$. In each area, it is determined how many times a subgroup of SSBs is found to be the strongest among all available SSBs. Those areas with the highest presence of the strongest SSB subset will be the ones under the influence of those beams. Figs. \ref{coverage_maps}(a)-(k) show density maps that illustrate the SSB influence area for different SSBs subsets. Red color denotes zones where a specific SSB subset is found to be the one with strongest SS-RSRP, while blue tones indicate areas where the SS-RSRP is not maximum among all SSBs. Figs.~\ref{coverage_maps}(a)-(c) groups the SSBs according to elevation/downtilt angle. Fig. \ref{coverage_maps}(a) considers those SSBs with $\theta = 0\degree$. These SSBs cover the farthest areas of the factory, even in the NLoS area that does not correspond angularly with the azimuth angle of the beams. This fact is explained by the reflections and scattering that occur in the ventilation ducts that are at the same height as the transmitter module, as shown in Fig. \ref{fotos}(b). This environment favors reflection and scattering phenomena covering the NLoS areas of the dense hall. As the downtilt is increased in Figs. \ref{coverage_maps}(b), $\theta = 7\degree$, and \ref{coverage_maps}(c), $\theta = 15\degree$, the coverage areas are concentrated around the location of the transmitter module. In fact, in the latter case, it is depicted that the beam with the higher downtilt only provides coverage to a small area just below the transmitter. Figs.~\ref{coverage_maps}(d)-(k) group the SSBs according to several azimuth angles, considering the two uppermost ($\theta = 0\degree$ and $\theta = 7\degree$) elevation/downtilt cases. Specifically, the four most south-facing SSBs, shown in Fig.~\ref{coverage_maps}(d), are the ones that cover the dense hall due to reflections and scattering in the ventilation ducts, following the same reasoning as in Fig.~\ref{coverage_maps}(a). As the beams are beamformed to cover northward areas, it is found that the areas under the influence of the beam also move towards the north part of the factory. The only exception is found in Fig.~\ref{coverage_maps}(k), where the beam points out directly at the only metal structure whose height is higher than the transmitter in the sparse hall. This causes the beam to be obstructed, reflecting off that structure, thus appearing the coverage area in the southern part of the sparse hall. In summary, beam-specific coverage areas show that the beam determination in LoS condition can be approximated by an a priori assignment due to the high correlation between the beam Direction-of-Departure and the regions with the maximum SS-RSRP.  This is due to the maximum contribution of LoS versus other multipath components, as previously predicted in Fig.~\ref{angular_analysis}. On the NLoS side, the density maps show that reflection and scattering phenomena at metallic interfaces allow radio propagation to areas outside the angular range of the beams. This fact can be exploited in network design and deployment to cover NLoS blocked by walls or large machinery.

As a last experiment, the beam switch-off problem is considered \cite{optimization}. It is desired to know how the coverage in the factory degrades if some of the 27 SSBs available in the transmitter are turned off (configuration B). If the degradation in terms of RSRP is not high, the selective shutdown of some beams can be considered an energy-saving solution, thus increasing the energy efficiency in the network. For this purpose, the following optimization problem is proposed:

\begin{equation}
\min \left(f(set) = \frac{1}{N}\sum_x \sum_y \overline{\textrm{RSRP}}_{\max , x, y}-\overline{\textrm{RSRP}}_{set, x, y}\right)
\end{equation}

\begin{equation}
\textrm{ subject to: } \quad \# SSB_{set} \leq \xi
\end{equation}

\noindent where $x$ and $y$ denotes the indices of the squared grid previously defined for Fig. \ref{RSRP_map}(c), $N$ is the number of grids, $\overline{\textrm{RSRP}}_{\max, x, y}$ stands for the average RSRP over all the measurements in the grid ($x,y$) when every beam is turned on. $\overline{\textrm{RSRP}}_{set, x, y}$ is the average RSRP over all the measurements in the grid ($x,y$) when a subgroup of SSBs specified in $set$ is turned on. $\# SSB_{set}$ is the number of SSBs turned on simultaneously, under the constraint that it must be less than the threshold $\xi$.

This problem has been solved by using the MI-LXPM genetic algorithm \cite{MI-LXPM}, which is able to deal with binary optimization problems. The space of feasible solutions is a binary vector with 27 elements, each corresponding to the on/off state of each SSB. The algorithm is run 10 times independently for three possible $\xi$ threshold values: 3, 5 and 10. In all 10 iterations, the same solution was found for each threshold. Figs. \ref{optimization}(a)-(c) show the SSB configuration found by the genetic algorithm for each case. In Fig. \ref{optimization}(a), when only three SSBs are turned on, the algorithm chooses all of them from the middle row. Two of them point to the south side of the factory, thus covering the dense hall, while one of them points to the north side. With this configuration, the central part of the sparse hall is uncovered, as it is illustrated by the high values of $f_{x,y}$ found in this area. On average, with 3 beams on, the average RSRP degrades by 3.8 dB. In Fig. \ref{optimization}(b), $\xi$ is raised to five, with the algorithm finding the same three beams as in the previous case, but adding two more in the central part. This solves the problem previously described where the sparse hall was uncovered by the beams, thus decreasing the average RSRP degradation to 1.9 dB. Finally, Fig. \ref{optimization}(c) shows the case where 10 SSBs are turned on. Observing the factory map, it can be concluded that the signal degradation is minimal, with an average degradation of 0.7 dB.

As an alternative, a second algorithm with lower computational complexity is proposed to solve the beam \mbox{switch-off} problem. DBSCAN \cite{DBSCAN} is an unsupervised learning algorithm within machine learning techniques that is able to discover data clusters with high density given a data space. This algorithm has been successfully employed to determine the set of beams that fits the user distribution in a certain deployment~\cite{DBSCAN_IA}. Focusing on the beam switch-off problem, it is possible to cluster the SSBs given the highest RSRP at a given position. Thus, those SSBs whose RSRP density is maximum in DBSCAN are the ones selected to be turned on. \mbox{Figs. \ref{optimization}(d)-(f)} show the SSB configuration found for the three previous threshold values. For $\xi$ = 3, DBSCAN opts for a different strategy than the genetic algorithm, mainly covering the central part of the sparse hall. Afterward, when $\xi$ increases, the beams covering the farthest parts of the TX module are turned on. Note that the average RSRP degradation is slightly higher than in the case of genetic algorithms. This fact is expected since the cost function (eq. 4) minimizes the degradation, whereas DBSCAN seeks SSBs whose RSRP density has historically been the highest, thereby disregarding the cost function. In return, DBSCAN low complexity determines the beams within a few seconds, while genetic algorithms can take up to a minute. Taking into account the better performance of genetic algorithms and the lower complexity of DBSCAN, it is possible to consider a combination of both techniques in a real deployment. Specifically, genetic algorithms can be applied at regular intervals to determine the SSB set turned on, while based on real-time network information, these SSBs can be updated following the data-driven approach proposed by DBSCAN due to its lower computational complexity.

In summary, the previous results indicate that an energy-saving solution based on beam switch-off is feasible due to the fact that the degradation of the average received power is minimal. It has been demonstrated that with 10 out of 27 beams on, the RSRP decreases on average by only 0.7 dB with respect to the case where all SSBs are on. Although it has to be taken into consideration that, in the case of applying this technique, the reliability of the network may decrease in terms of beam failure due to the lower availability of beams, as shown in Section III.B.

\section{Conclusion}

This work presents two measurement campaigns performed in the 5G Smart Production Lab, which resembles a small factory. The analysis of both campaigns has been carried out in a 5G NR network at 26 GHz in the FR2 frequency range under operational conditions. These measurement campaigns have been carried out with AMRs which are configured to self-navigate the industrial lab, following predefined routes. The analysis of the RSRP received by an AMR in terms of the configuration of a transmitter module has been carried out, allowing the generation of coverage maps throughout the factory.

The experimental characterization based on the comparison of both campaigns proves: (i) a higher separation of the beams in the TX (configuration B) provides better overall coverage in the factory. In exchange, the higher separation decreases the average power of the best backup beam, worsening the quality of the beam recovery procedure. Otherwise, a set of angularly closer beams (configuration A), obtains a backup channel with higher RSRP. (ii) Path gain fitting of the factory measurements for both campaigns agrees with models developed at these frequencies for industrial environments in both LoS and NLoS condition, which validates previous models for radio planning in factory-based scenarios. (iii) For an RSRP target above -100 dBm, cell radii up to 26 meters can be considered with 99.8\% coverage. In NLoS condition, the coverage decreases to 70.2\% (configuration A) and 82.7\% (configuration B) for 23-meter cell radii. 

Concerning the beam management procedure analysis, the following conclusions are drawn: (i) for beam sweeping and switching analysis, the beam angular coverage distribution in the LoS region can be approximated by that expected in free space, which allows an a priori allocation of resources based on the AMR position. (ii) The beam recovery analysis shows that an alternative serving beam is available with power losses between 0.7 dB and 1.6 dB on average when all beams are available. TX beam distributions with high spatial density decrease the power gap between the serving and the backup beam. (iii) The development of coverage maps based on subsets of beams demonstrates that reflection and scattering propagation mechanisms in the factory ventilation ducts allow the signal to propagate to NLoS areas of the factory. (iv) A beam switch-off problem optimization indicates the feasibility of beam switch-off as an energy-saving solution with average RSRP losses of 0.7 dB when 17 of the 27 available beams are switched off (63\%).

The results detailed in this study, which consider propagation, mobility and beam management aspects, offer valuable insights for the radio planning of similar industrial hall scenarios using FR2 wall-mounted directional antennas. Given the increasing significance of Industry 4.0 and wireless automation, thorough analyses are not only required from an empirical and experimental perspective but also from theoretical approaches. Future research directions include the development of geometrical models, based on deterministic techniques such as ray-tracing, and statistical models through comprehensive 3D spatial modeling. Moreover, considering the experimentation conducted in this study, an in-depth analysis of specific factors that may degrade communication performance, such as human blockage or beam misalignment, is crucial for efficient deployments. With the emergence of machine learning techniques, these are suggested as promising candidates for developing methods and algorithms for beam management.

\section*{Acknowledgments}

The authors would like to thank D.~Chizhik, J.~Du and R.~A.~Valenzuela, from Nokia Bell Labs, for their constructive comments.

\end{document}